\documentclass{article}
\usepackage{authblk}
\usepackage{geometry}
\usepackage[titletoc]{appendix}
\usepackage{enumitem}
\usepackage{booktabs,diagbox}
\usepackage{hyperref}
\usepackage{xcolor}

\usepackage{tikz} 
\usetikzlibrary{cd}
\usetikzlibrary{arrows}

\usepackage{amssymb,amsmath,amsthm,mathtools,slashed,bm} 
\def\-{\raisebox{.75pt}{-}}

\numberwithin{table}{section}
\numberwithin{equation}{section}

\theoremstyle{definition}
\newtheorem{defn}{Definition}[section]

\newtheorem{exmp}{Example}[section]
\newtheorem{rmk}{Remark}[section]
\theoremstyle{plain}
\newtheorem{lem}{Lemma}[section]
\newtheorem{prop}{Proposition}[section]
\newtheorem{thm}{Theorem}[section]
\newtheorem*{thm*}{Theorem}
\newtheorem{cor}{Corollary}[section]

\usepackage{amsrefs}

\title{Mathematical Structures of Cohomological Field Theories}
\author[1]{Shuhan Jiang}
\date{}
\affil[1]{Max Planck Institute for Mathematics in the Sciences, 04103 Leipzig}

\begin{document}
	
	\maketitle
	
	\begin{abstract}
		A mathematical framework of cohomological field theories (CohFTs) is formulated in the language of bigraded manifolds. Algebraic properties of operators in CohFTs are studied. Methods of constructing CohFTs, with or without gauge symmetries, are discussed. In particular, a generalization of the Mathai-Quillen formalism is given. Examples such as topological quantum mechanics, topological sigma model, topological M-theory, and topological Yang-Mills theory can be obtained uniformly using this new formalism.  
	\end{abstract}

\section{Introduction}

A cohomological field theory (CohFT) is a Lagrangian field theory which possesses a scalar supersymmetry $Q$ with $Q^2=0$ \cites{Witten1988,Witten1991}. The physical operators of interest in such a theory are solutions to the following equations
\begin{align}
	Q \mathcal{O}^{(p)} = d \mathcal{O}^{(p-1)} \label{eq:desc} 		
\end{align}
for $1 \leq p \leq n$ with $Q \mathcal{O}^{(0)} = 0$, where $d$ is the de Rham differential of an $n$-dimensional Riemannian manifold $M$. The expectation value of $\int_{\gamma_p}\mathcal{O}^{(p)}$, where $\gamma_p$ is a $p$-cycle in $M$, does not depend the Riemannian metric. That is to say, it can be seen as a smooth invariant of $M$. Many famous invariants in mathematics, e.g., the Donaldson invariants, Gromov-Witten invariants, and Seiberg-Witten invariants, can be obtained in this way. 

There exists various mathematical approaches to CohFTs based on the BRST cohomology \cite{Baulieu1988,Baulieu1989}, the equivariant cohomology \cite{Atiyah1990}, and the master equation in the BV formalism \cite{Alexandrov1997}. However, most of these approaches focus only on the construction of Lagrangians, lacking a careful treatment of the algebraic structures of operators in CohFTs. It is the goal of this paper to build a new mathematical framework for CohFTs systematizing the previous ones, within which a complete classification of the solutions to \eqref{eq:desc} is available.
	
The first step toward such a framework is to generalize the theory of supermanifolds (or $\mathbb{Z}_2$-graded manifolds) to a theory with richer grading structures. Such a theory was developed by the author in \cite{Jiang2022}, where $\mathbb{Z}_2$ is replaced by a cancellative commutative semiring $\mathcal{I}$ to yield the notion of a monoidally graded manifold. Here we are mainly interested in the case of $\mathcal{I}=\mathbb{Z} \times \mathbb{Z}$, i.e., bigraded manifolds.  The operator $\mathcal{O}^{(p)}$ can be viewed (classically) as a function of degree $(p,n-p)$ over a bigraded manifold. It follows that $Q$ and $d$ should be viewed as vector fields of degrees $(0,1)$ and $(1,0)$, respectively. Moreover, the supersymmetry algebra can be extended to include a vector supersymmetry $K$, a vector field of degree $(1,-1)$. $K$ together with $Q$ and $d$ satisfies the following relations
\[
Q^2=0, \quad QK+KQ=d, \quad Kd+dK=0.
\]
We call the corresponding extended supersymmetry algebra as the $QK$-algebra. $K$ can be used to produce particular solutions to \eqref{eq:desc} called $K$-sequences by setting $\mathcal{O}^{(p)}=\sum_{q=0}^p \frac{1}{(p-q)!}K^{p-q} \mathcal{W}^{(q)}$, where $\mathcal{W}^{(0)}=\mathcal{O}^{(0)}$ and $\mathcal{W}^{(q)}$ is any (non-exact) $Q$-closed function of degree $(q,n-q)$ for $1 \leq q  \leq n$. The main result of this paper is
\begin{thm*}
	Every solution to \eqref{eq:desc} is cohomologically a $K$-sequence.
\end{thm*}

The paper is organized as follows: In Section 2, knowledge about equivariant cohomology is reviewed to motivate mathematical constructions later. In Sections 3 and 4, we build the geometric settings for CohFTs and prove the main theorem. In Section 5, we investigate Witten's idea of topological twistings \cite{Witten1988} and show that the twisting of a super Poincar\'e algebra gives naturally rise to a $QK$-algebra. In Section 6, we provide a systematic formalism of constructing CohFTs as a generalization of the methods in \cite{Witten1988,Witten1991,Baulieu1988,Baulieu1989,Ouvry1989,Atiyah1990,Birmingham1991,Kalkman1993,Blau1993}. In this new formalism, one can create CohFTs by specifying just a small collection of data, as shown in Section 7. 
	
\section{Equivariant Cohomology}
	
In this section, we briefly describe three different approaches to equivariant cohomology, namely, the Weil model, the Cartan model, and the Kalkman model. We use the last one to reformulate the Mathai-Quillen construction of the Thom class.
	
Naturally, one should expect that the equivariant cohomology of a $G$-space $X$ tells us both the topological information of $X$ and the information about the $G$-action on $X$. A naive choice is the cohomology of the quotient space $X/G$. This is not right, since $X/G$ remembers nothing about the stabilizer of the group action at each point $x \in X$ unless the group action is free. The key idea here is to consider the Cartesian product of a ``universal $G$-space'' $EG$ and $X$ where
\begin{enumerate}
	\item $EG$ is contractible, hence it does not provide any new topological information;
	\item $EG$ has a free $G$-action. (It follows that $G$ acts also freely on $EG \times X$.)
\end{enumerate}     
\begin{defn}
	A contractible space $EG$ with a free $G$-action is called a universal $G$-space. Let $X$ be a $G$-space. The quotient space $X_G=(EG \times X)/G$ is called a homotopy orbit space of $X$. In particular, the homotopy orbit space of a one-point space is called a classifying space for $G$, denoted by $BG$.
\end{defn}
A standard result is that $EG$ always exists for a topological Lie group $G$. Moreover, all homotopy orbit spaces of a $G$-space $X$ are homotopic equivalent to each other. 
\begin{defn}\label{defofeqco}
	The equivariant cohomology of a $G$-space $X$, denoted by $H_G(X)$, is defined as the singular cohomology of the homotopy orbit space $X_G$ of $X$.
\end{defn}
Let's switch to the smooth category and assume $G$ to be compact. By the quotient manifold theorem, the homotopy orbit space $X_G$ is also smooth and we can replace the singular cohomology of it in Def \ref{defofeqco} by its de Rham cohomology.
\begin{rmk}
	The universal $G$-space $EG$ is often obtained as a direct limit of some direct system of finite dimensional manifolds. The de Rham complex of $EG$ can be then defined as the inverse limit of the induced inverse system of de Rham complexes over such finite dimensional manifolds.
\end{rmk}
Let $G$ be a Lie group with Lie algebra $\mathfrak{g}$. To each $\xi \in \mathfrak{g}$ we can associate a vector field $v_{\xi}$ on a $G$-manifold, which again induces a contraction $\iota_{\xi}$ and a Lie derivative $\mathrm{Lie}_{\xi}$ on the de Rham complex of the $G$-manifold. Let $d$ denote the de Rham differential. Fix a basis $\{\xi_a\}$ of $\mathfrak{g}$. Let $\iota_a$ and $\mathrm{Lie}_a$ denote the contraction and Lie derivative associated to $\xi_a$. $d$, $\iota_a$, $\mathrm{Lie}_a$ satisfy the following relations
\begin{align}
	&[\mathrm{Lie}_a,\mathrm{Lie}_b] = f^c_{ab} \mathrm{Lie}_c, \quad [\mathrm{Lie}_a,\iota_b] = f^c_{ab} \iota_c, \quad
	[\mathrm{Lie}_a,d] = 0, \label{e_5_1a} \\
	&\{d,d\}=0, \quad 
	\{\iota_a,\iota_b\}=0, \quad 
	\{d,\iota_a\}=\mathrm{Lie}_a. \label{e_5_1b}    
\end{align}
\begin{rmk}
	These relations are not independent from each other. It is an easy exercise to show that $[\mathrm{Lie}_a,\mathrm{Lie}_b] = f^c_{ab} \mathrm{Lie}_c$ and $[\mathrm{Lie}_a,d] = 0$ can be derived from the others.
\end{rmk}     
Recall that a Lie superalgebra $L=L_{even} \oplus L_{odd}$ is specified by \cite{Leites1980}
\begin{enumerate}
	\item a Lie algebra $L_{even}$;
	\item an $L_{even}$-module $L_{odd}$ and a bilinear pairing $[\cdot,\cdot]$ such that $[x,\cdot]$ is the action of $x \in L_{even}$ on $L_{odd}$ and $[\cdot,x]: =-[x,\cdot]$;
	\item a symmetric bilinear paring $\{\cdot,\cdot\}: L_{odd} \times L_{odd} \rightarrow L_{even}$ that is a homomorphism of $L_{even}$-modules and satisfies the Jacobi identity
	\begin{align*}
		\{x,\{y,z\}\} + \{y,\{z,x\}\} + \{z,\{x,y\}\}=0
	\end{align*}
	for $x,y,z \in L_{odd}$. 
\end{enumerate}
In this case, $L_{even}$ is spanned by $\mathrm{Lie}_a$. $L_{odd}$ is spanned by $d$ and $\iota_a$. In other words, we have $L_{even}=\mathfrak{g}$ and $L_{odd} = \mathbb{R} \oplus \mathfrak{g}$. Elements of $L_{even}$ act on $\mathbb{R} \subset L_{odd}$ trivially and act on $\mathfrak{g} \subset L_{odd}$ via the adjoint representation. The symmetric pairing is given by (\ref{e_5_1b}). Moreover, $L$ is a Lie graded algebra by assigning degrees $0$, $1$ and $-1$ to $\mathrm{Lie}_a$, $d$ and $\iota_a$, respectively. 
\begin{defn}
	Let $W$ be an $L$-module. Let $\mathrm{Lie}_a$, $d$, $\iota_a$ be a basis of $L$. An element $\alpha \in W$ is called horizontal if $\iota_a \alpha =0$. A horizontal element $\alpha$ is called basic if in addition $\mathrm{Lie}_a \alpha =0$. 
\end{defn} 
Let $W_{bas}$ denote the submodule of basic elements in $W$. Note that for $\alpha \in W_{bas}$, $d\alpha$ is also in $W_{bas}$. As an alternative approach to equivariant cohomology, one can define the equivariant cohomology of a $G$-manifold $X$ as the cohomology of the basic submodule $(\Omega(EG) \otimes \Omega(X))_{bas}$ of the $L$-module $\Omega(EG) \otimes \Omega(X)$. In practice, $\Omega(EG)$ is too large to work with. It is shown in \cite{Guillemin2013} that one can replace it with the following algebra.
\begin{defn}\label{d_5_1}
	The Weil Algebra is the commutative graded algebra
	\begin{align*}
		W(\mathfrak{g})= \Lambda(\mathfrak{g}^*) \otimes \mathrm{S}(\mathfrak{g}^*)
	\end{align*}
	where elements of $\mathfrak{g}^* \subset \Lambda(\mathfrak{g}^*)$ have degree $1$ and elements of $\mathfrak{g}^* \subset \mathrm{S}(\mathfrak{g}^*)$ have degree $2$.
\end{defn}
Let's show that $W(\mathfrak{g})$ has an $L$-module structure. Let $\xi^a$ be a basis of $\mathfrak{g}^*$, which yields generators $\{\theta^a\}$ and $\{\phi^a\}$ for $\Lambda(\mathfrak{g}^*)$ and $\mathrm{S}(\mathfrak{g}^*)$, respectively. We can now define $\iota_a$ and $d$ by setting
\begin{align}
	\iota_a\theta^b &= \delta_{a}^b,\quad \iota_a \phi^b=0, \label{e_5_5} \\
	d\theta^a &= \phi^a - \frac{1}{2}f^a_{bc}\theta^b\theta^c,\quad d\phi^a=f^a_{bc}\phi^b\theta^c. \label{e_5_6}
\end{align}
Consequently, we have
\begin{align}
	\mathrm{Lie}_a \theta^b=-f^b_{ac}\theta^c,\quad \mathrm{Lie}_a \phi^b=-f^b_{ac}\phi^c.
\end{align}
It is an easy exercise to verify that $\mathrm{Lie}_a$, $d$, $\iota_a$ satisfy (\ref{e_5_1a}) and (\ref{e_5_1b}).  Moreover, one can show that $W(\mathfrak{g})$ is acyclic, which is not surprising because $\Omega(EG)$ is also acyclic.

\begin{defn} 
	The universal connection and curvature of the Weil algebra are
	\begin{align*}
		\theta = \theta^a \otimes \xi_a \in W^1(\mathfrak{g})\otimes \mathfrak{g}, \quad
		\phi = \phi^a \otimes \xi_a \in W^2(\mathfrak{g})\otimes \mathfrak{g}.
	\end{align*}
\end{defn}

Let $P$ be a principal $G$-bundle. $\Omega(P)$ is an $L$-module. The connection and curvature forms on $P$ determine maps
\begin{align*}
	\mathfrak{g}^* \rightarrow \Omega^1(P),\quad \mathfrak{g}^* \rightarrow \Omega^2(P)
\end{align*}
which induce a homomorphism of commutative graded algebras
\begin{align*}
	\phi_W: W(\mathfrak{g})\rightarrow \Omega(P)
\end{align*}
known as the Weil homomorphism.
\begin{thm}[\cite{Mathai1986}]
	$\phi_W$ is a morphism of $L$-modules. Moreover, it sends the universal connection and curvature of $W(\mathfrak{g})$ to the connection and curvature forms on $P$.
\end{thm}
Now, let's consider the tensor product $W(\mathfrak{g})\otimes \Omega(X)$. It has an $L$-module structure by defining the contractions, the Lie derivatives and the differential to be
\begin{align*}
	\iota_a \otimes 1 + 1 \otimes \iota_a,\quad \mathrm{Lie}_a \otimes 1 + 1 \otimes \mathrm{Lie}_a,\quad d \otimes 1 + 1 \otimes d.
\end{align*}
Let $\Omega_G(X)$ denote the basic part of $W(\mathfrak{g})\otimes \Omega(X)$. Let $d_W$ denote the differential on $\Omega_G(X)$ induced from the differential on $W(\mathfrak{g})\otimes \Omega(X)$.
\begin{defn}
	$(\Omega_G(X),d_W)$ is called the Weil Model for the equivariant cohomology of a $G$-manifold $X$. $d_W$ is called the Weil differential.
\end{defn}
Recall that differential forms on $P \times_G X$ can be identified with basic forms on $P \times X$. The Weil homomorphism induces a new homomorphism $\phi_{CW}$ of commutative graded algebras through the commutative diagram 
\[ \begin{tikzcd}
	W(\mathfrak{g}) \otimes \Omega(X) \arrow{r}{\phi_W} &\Omega(P \times X) \\
	\Omega_G(X) \arrow{r}{\phi_{CW}} \arrow[hook]{u}& \Omega(P \times_G X) \arrow[hook]{u}
\end{tikzcd} \]   
$\phi_{CW}$ is known as the Chern-Weil homomorphism. It induces a homomorphism of cohomologies
\begin{align*}
	H_G^*(X) \rightarrow H^*(P \times_G X)
\end{align*}
which does not depend on the choice of connection on $P$. In particular, the Chern-Weil homomorphism induces a homomorphism $S(\mathfrak{g}^*)^G \rightarrow H(P/G)$ because $H_G(\mathrm{pt}) \cong S(\mathfrak{g}^*)^G$. 

\begin{defn}
	The Mathai-Quillen map is the automorphism map $j = \exp{(-\theta^a \otimes \iota_a)}$ of $W(\mathrm{g})\otimes \Omega(X)$. The differential $d_K = j \circ d_W \circ j^{-1}$ is called the Kalkman differential. $(W(\mathrm{g})\otimes \Omega(X), d_K)$ is called the Kalkman model of the equivariant cohomology of a $G$-manifold $X$.
\end{defn}
One can show that 
\begin{align*}
	d_K = d \otimes 1 + 1 \otimes d + \theta^a \otimes \mathrm{Lie}_a - \phi^a \otimes \iota_a,
\end{align*}
and that $\iota_a \otimes 1 = j \circ (\iota_a \otimes 1 + 1 \otimes \iota_a) \circ j^{-1}$, $(\mathrm{Lie}_a \otimes 1 + 1 \otimes \mathrm{Lie}_a) = j \circ (\mathrm{Lie}_a \otimes 1 + 1 \otimes \mathrm{Lie}_a) \circ j^{-1}$. Consequently, the basic part of $W(\mathrm{g})\otimes \Omega(X)$ in the Kalkman model of $X$ is just $(S(\mathfrak{g}^*) \otimes \Omega(X))^G$, and the restriction of the Kalkman differential to $(S(\mathfrak{g}^*) \otimes \Omega(X))^G$, denoted by $d_C$, takes the form $d_C = d \otimes 1 - \phi^a \otimes \iota_a$.
\begin{defn}
	$((S(\mathfrak{g}^*) \otimes \Omega(X))^G, d_C)$ is called the Cartan Model of a $G$-manifold $X$. $d_C$ is called the Cartan differential.
\end{defn}

Let $G=SO(n)$, $n=2m$. Let $\rho$ be the standard representation of $G$ on $V=\mathbb{R}^n$. We also use $V$ to denote the $n$-dimensional translation group and its Lie algebra. Let $\theta$ and $\phi$ be the universal connection and curvature of the Weil algebra of $G$. Let $w^i$ and $b_i$ denote the coordinate functions of $V$ and $V^*$ respectively. Let $\chi_i$ denote the odd coordinate functions of $\Pi V^*$. In 
\cite{Mathai1986}, Mathai and Quillen defined the following element\footnote{The imaginary unit $i$ is introduced so that we will get an integrable Gaussian function of $x$ after integrating out the ``auxiliary" field $b$.}     
\begin{align*}
	U=(2\pi)^{-n}\int  d \chi d b \exp\left(-b^t(b/2 + i w) + \frac{1}{2}\chi^t \phi \chi + i (dw + \theta w)^t \chi \right)
\end{align*}
of degree $n$ in $\Omega_G(V)$, where $\phi \chi = \phi^a \otimes \rho(\xi_a) \chi$, $\theta w = \theta^a \otimes \rho(\xi_a) w$, $\int d \chi d b$ is the Berezin integral over the even variables $b_i$ and the odd variables $\chi_i$.
\begin{prop}\label{mqu}
	$U$ is closed in $\Omega_G(V)$.
\end{prop}
To prove this proposition, let 
\begin{align}\label{Lunifin}
	L = b^t(b/2 + i w) - \frac{1}{2}\chi^t \phi \chi - i \chi^t (dw + \theta w).
\end{align}
$L$ can be seen as an element in $W(\mathfrak{g}) \otimes \Omega(V^*) \otimes \Omega(V)$. Let $G \ltimes V$ be the semi-direct product of $G$ and $V$ induced by $\rho$. Let $G \ltimes V$ act on $V$ through $\rho$ only, i.e., we require that the translation part of the group acts trivially. Let $\mathfrak{g} \ltimes V$ denote its Lie algebra. With a slight abuse of notation, we use $\Omega(V^*)$ to denote $\mathrm{S}(V^*) \otimes \Lambda(V^*)$. Note that $W(\mathfrak{g} \ltimes V)$ is isomorphic to $W(\mathfrak{g}) \otimes \Omega(V^*)$ as commutative graded algebras. Moreover,
\begin{lem}
	The Weil model $(W(\mathfrak{g} \ltimes V) \otimes \Omega(V),d_W)$ is isomorphic to the differential graded algebra $(W(\mathfrak{g}) \otimes \Omega(V^*) \otimes \Omega(V),s)$. The differential $s$ takes the form
	\begin{align*}
		s = d_K \otimes 1 + 1 \otimes d,
	\end{align*}
	where $d_K = d \otimes 1  + 1 \otimes \delta_K  + \theta^a \otimes \mathrm{Lie}_a - \phi^a \otimes \iota_a$ is the Kalkman differential of $W(\mathfrak{g}) \otimes \Omega(V^*)$, with $\delta_K$ denoting the Koszul differential on $\Omega(V^*)$. 
\end{lem}
\begin{proof}
	We need to show that the differential $d$ on $W(\mathfrak{g} \ltimes V)$ is equivalent to the Kalkman differential $d_K$ on $W(\mathfrak{g}) \otimes \Omega(V^*)$. Let $\xi_a$ and $t^i$ be bases of $\mathfrak{g}$ and $V$, respectively. We can write $\rho(\xi_a) t^i = \rho_{aj}^i t^j$. The Lie bracket of $\mathfrak{g} \ltimes V$ is given by
	\begin{align*}
		[(\xi_a, t^i),(\xi_b, t^j)] = ([\xi_a, \xi_b], \rho(\xi_a)t^j - \rho(\xi_b)t^i) = (f^c_{ab} \xi_c, \rho_{ak}^j t^k - \rho_{bk}^i t^k).
	\end{align*}
	It then follows from (\ref{e_5_6}) that
	\begin{align*}
		&d \theta^a = \phi^a - \frac{1}{2}f^a_{bc}\theta^b \theta^c, \quad d \phi^a = -f^a_{bc}\theta^b \phi^c, \\
		&d \chi^i = b^i -\rho^i_{aj}\theta^a\chi^j,\quad d b^i = \rho^i_{aj} \phi^a \chi^j  - \rho^i_{aj} \theta^a b^j.
	\end{align*}
	On the other hand, one easily can check that 
	\begin{align*}
		&\mathrm{Lie}_a b^i = -\rho_{aj}^i b^j, \quad \mathrm{Lie}_a \chi^i = -\rho_{aj}^i\chi^j, \\
		&\iota_a b^i = -\rho_{aj}^i \chi^j, \quad \iota_a \chi^i = 0. 
	\end{align*}
	The rest of the proof is straightforward.
\end{proof}
$W(\mathfrak{g} \ltimes V) \otimes \Omega(V)$, or equivalently, $W(\mathfrak{g}) \otimes \Omega(V^*) \otimes \Omega(V)$, is also an $L$-module by setting
\begin{align*}
	\mathrm{Lie}_a = \left(\mathrm{Lie}_a \otimes 1 + 1 \otimes \mathrm{Lie}_a \right) \otimes 1 + 1 \otimes 1 \otimes \mathrm{Lie}_a, \quad \iota_a = (\iota_a \otimes 1) \otimes 1 + 1 \otimes 1 \otimes \iota_a.
\end{align*}
For simplicity, we omit the indices of the coordinate functions. The action of $s$ on coordinate functions can then be written as
\begin{align}
	&s \theta = \phi - \theta\theta, \quad s \phi = [\phi,\theta], \label{smq0} \\
	&s w = dw,\quad s b = -\theta b + \phi \chi, \label{smq1}\\
	&s dw = 0,\quad s \chi = b - \theta \chi. \label{smq2}
\end{align} 
\begin{rmk}
	It is always fun to check $s^2=0$ by direct computations. The non-trivial ones are
	$
	s^2 b = -(s\theta)b + \theta (sb) + (s\phi)\chi + \phi (s\chi)
	=-(\phi-\theta \theta)b + \theta (-\theta b + \phi \chi) + ([\phi,\theta])\chi + \phi (b - \theta \chi) 
	= 0
	$, and
	$
	s^2 \chi = sb -s(\theta)\chi + \theta (s\chi)
	= (-\theta b + \phi \chi) - (\phi-\theta \theta)\chi + \theta(b - \theta \chi)
	= 0
	$.
\end{rmk}
\begin{lem}
	$L$ is exact in $(W(\mathfrak{g} \ltimes V) \otimes \Omega(V))_{bas}$.
\end{lem}
\begin{proof}
	The exactness of $L$ follows from direct computations.
	\begin{align*}
		s \left(\chi^t (i w + b/2)\right) &= (b-\theta \chi)^t(iw+b/2) - \chi^t(idw+(-\theta b + \phi \chi)) \\
		&= b^t(iw+b/2) - (-\chi^t\theta^t)(iw+b/2)- \chi^t(idw -\theta b/2) - \frac{1}{2} \chi^t \phi \chi \\
		&= b^t(iw+b/2) - \frac{1}{2} \chi^t \phi \chi - \chi^t((i\theta w + \theta b/2) + (idw -\theta b/2)) \\
		&= b^t(iw+b/2) - \frac{1}{2} \chi^t \phi \chi - \chi^t i(dw + \theta w).
	\end{align*}
	We use the skew-symmetric property of $\theta$ in the third step. $L$ is basic because $\chi^t (i w + b/2)$ is $\mathfrak{g}$-invariant and does not contain $\theta$ and $dw$.
\end{proof}
Let $\mathcal{S}(V^*)$ denote the space of Schwartz functions over $V^*$. For every $\alpha \in \mathcal{S}(V^*) \otimes \Lambda(V^*)$, the super Fourier transform $\mathcal{F}$ of $\alpha$ is defined by
\begin{align*}
	\mathcal{F}(\alpha) = \int_{V^*} \alpha \exp(-i(b^t w - \chi^t dw)) \in \Omega(V).
\end{align*}
It is not hard to show that $\mathcal{F}$ commutes with both $d$ and $\iota_a$ \cite{Kalkman1993}, hence also $\mathrm{Lie}_a$.

Now, apply the Mathai-Quillen map $j=\exp(-\theta^a \otimes \iota_a)$ to $W(\mathfrak{g}) \otimes \Omega(V^*)$. This is equivalent to a change of coordinates which sends $b$ to $b - \theta \chi$. In the new coordinates, we have
$
s = d \otimes 1 \otimes 1 + 1 \otimes \delta_K \otimes 1 + 1 \otimes 1 \otimes d
$, and 
$
L = s \alpha + i(b^t w - \chi^t dw)
$, where
$
\alpha = \chi^t (b-\theta \chi)/2
$. 
It is easy to see that $\alpha$ is a basic element in $W(\mathfrak{g}) \otimes \Omega(V^*)$. Moreover, the element $\exp (-s \alpha)$ is a closed basic element in $W(\mathfrak{g}) \otimes \Omega(V^*)_{S}$ due to the Gaussian factor $\exp(-b^tb/2)$. Proposition \ref{mqu} is then proved by observing that
\begin{align*}
	\int d\chi db \exp(-L) = \mathcal{F}(\exp(-s\alpha)).
\end{align*}

Note that integrating out $b$ and $\chi$ will give us a factor $(2\pi)^m$. The component of $U$ with top de Rham degree is $(2\pi)^{-m}\exp(-\frac{1}{2}w^2) dw^1 \dots dw^n$. It follows that
$
\int_V U = 1
$.
Let $P$ be a principal $G$-bundle over a manifold $\Sigma$. Let $A$ be a connection $1$-form on $P$. Let $E$ be an associated vector bundle to $P$ of rank $2m$ equipped with a metric $(\cdot,\cdot)$ and a metric connection $\nabla$ induced by $A$. 
\begin{thm}[\cite{Mathai1986}]
	$U$ is a universal Thom form in the sense that for any such $E$, the Chern-Weil homomorphism $\phi_{CW}$ sends $U$ to a form representing the Thom class of $E$.
\end{thm}

\begin{rmk}\label{diffutc}
	We can also consider the Kalkman model of $W(\mathfrak{g}\ltimes V) \otimes \Omega(V)$. The differential $s$ is locally given by 
	\begin{align}
		&s w = dw - \theta w,\quad s b = -\theta b + \phi \chi, \label{smq5}\\
		&s dw = -\theta dw + \phi w, \quad s \chi = b - \theta \chi. \label{smq6}
	\end{align}
	This is not the right choice for constructing a universal Thom class, as $s$ restricted to $\Omega(V)$ is not the de Rham differential. However, we will need this kind of differentials later to construct cohomological field theories with gauge symmetries.
\end{rmk}

Let $v$ be a section of $E$, we can also obtain a representative for the Euler class $e_{\nabla}(E)$ of $E$ by setting
\begin{align*}
	e_{\nabla}(E)=v^*\phi_{CW}(U)=(2\pi)^{-n}\int d \chi d b \exp\left( -L \right),
\end{align*}
where $L =(b,b/2+iv) - \frac{1}{2}(\chi, R \chi) - i \chi(\nabla v)$. Let $x^{\mu}$ denote the coordinate functions of $\Sigma$. Identifying $dx^{\mu}$ with the odd variables $\eta^{\mu}$, we have
\begin{align}\label{0qm}
	L = (b,b/2 + iv) - i \chi(\nabla_{\eta} v) - \frac{1}{4}(\chi,R(\eta,\eta)\chi).
\end{align}
Let's take $E$ to be $T\Sigma$, $(\cdot,\cdot)$ to be a Riemannian metric $g$, and $\nabla$ to be the Levi-Civita connection determined by $g$. We have
\begin{align}
	&s x^{\mu} = \eta^{\mu},\quad s b_{\mu} = \Gamma^{\nu}_{\rho \mu}\eta^{\rho}b_{\nu} - \frac{1}{2}R^{\nu}_{\mu \rho \sigma} \eta^{\rho}\eta^{\sigma} \chi_{\nu} 
	\label{0qm1} \\
	&s \eta^{\mu}=0,\quad s \chi_{\mu}=b_{\mu} + \Gamma^{\nu}_{\rho \mu} \eta^{\rho} \chi_{\nu}, \label{0qm2}
\end{align}
where $\Gamma^{\nu}_{\rho \mu}$ is the Christoffel symbol and $R^{\nu}_{\mu \rho \sigma}$ is the Riemann curvature tensor. (\ref{0qm}) together with the supersymmetry transformations (\ref{0qm1}) and (\ref{0qm2}) give us a $0$-dimensional supersymmetric theory \cite{Blau1993}.   
\begin{rmk}
	The Mathai-Quillen map $j$ of $W(\mathfrak{g})\otimes \Omega(V^*)$ induces a change of coordinates, namely,
	\begin{align*}
		b_{\mu} \rightarrow b_{\mu} + \Gamma^{\nu}_{\rho \mu} \eta^{\rho} \chi_{\nu}.
	\end{align*}
	The differential $s$ in the new coordinates takes the form
	\begin{align*}
		&s x^{\mu} = \eta^{\mu},\quad s b_{\mu} = 0,\\
		&s \eta^{\mu}=0,\quad s \chi_{\mu}=b_{\mu},
	\end{align*}
	which is exactly the BRST differential $s$ appearing in \cite{Baulieu1989}.	The price one pays for this simplification is that $L$ will no longer be covariant unless $b$ is integrated out.
\end{rmk}

\section{Graded and Bigraded Manifolds}  

Let $\mathcal{I}$ be a cancellative commutative semiring. Let $V=\bigoplus_{i \in \mathcal{I}} V_i$ be an $\mathcal{I}$-graded vector space with $\dim V_i = m_i$. The symmetric algebra $\mathrm{S}(V)$ is the quotient algebra of the tensor algebra $\mathrm{T}(V)$ by the two-sided ideal generated by
\begin{align}\label{grasign}
	v \otimes w - (-1)^{p(d(v)d(w))}w \otimes v,
\end{align}
where we use $d(u)$ to denote the degree of $u \in V$, $p:\mathcal{I} \rightarrow \mathbb{Z}_2$ is the parity function, $(-1)^{(\cdot)}$ is the sign function of $\mathbb{Z}_2$ which sends $0 \in \mathbb{Z}_2$ to $1$ and $1 \in \mathbb{Z}_2$ to $-1$.
\begin{rmk}
Here we have to be careful about the sign factor appearing on the right hand side of (\ref{grasign}). Although both $\mathcal{I}$ and $\mathbb{Z}_2$ are semi-rings, $p$ is not necessarily a semi-ring homomorphism and we do not have $p(d(v)d(w)) = p(d(v))p(d(w))$ in general. Bearing this in mind, we will use the more familiar notation $p(v)p(w)$ to replace $p(d(v)d(w))$ from now on.   
\end{rmk}
\begin{rmk}
	From now on, we will often use 
	\begin{align*}
		[v,w]=vw - (-1)^{p(d(v)d(w))}wv
	\end{align*}
    to denote the commutator bracket of $v$ and $w$ in an $\mathcal{I}$-graded algebra.
\end{rmk}
\begin{defn}\label{imfd}
	An $\mathcal{I}$-graded manifold $\mathcal{M}$ of dimension $n|(m_i)_{i \in \mathcal{I}}$ is a ringed space $(M,\mathcal{O})$ where $M$ is an $n$-dimensional manifold and $\mathcal{O}$ is a sheaf of commutative $\mathcal{I}$-graded algebras over $M$ which is locally isomorphic to $C^{\infty}(\mathbb{R}^n) \otimes \mathrm{S}(V)$.
\end{defn}

\begin{rmk}
	In fact, one should consider the algebra of formal power series on $V$ instead of $\mathrm{S}(V)$ in Definition \ref{imfd} \cite{Jiang2022}. Since this subtlety will not infect our discussion later, we stick with the $\mathrm{S}(V)$ version for simplicity.
\end{rmk}

An $\mathcal{I}$-graded vector bundle $\pi: E \rightarrow M$ is a vector bundle such that the local trivialization map
$
\varphi_U: \pi^{-1}(U) \rightarrow U \times V
$
is a morphism of $\mathcal{I}$-graded vector spaces when restricted to $\pi^{-1}(x)$, $x \in U \subset M$. In other words, $E = \bigoplus_{k \in \mathcal{I}} E_k$ where $E_k$ are vector bundles whose fibers consist of elements of degree $k$. To any $\mathcal{I}$-graded vector bundle $E$ we can associate an $\mathcal{I}$-graded manifold $E[l]$ whose structure sheaf is defined as the sheaf of sections of $\mathrm{S}(\bigoplus_{k \in \mathcal{I}} (E_{k+l})^*)$, $l \in \mathcal{I}$. It is shown in \cite{Jiang2022} that every $\mathcal{I}$-graded manifold can be obtained (non-canonically) in this way.

\begin{rmk}
	It is also possible to generalize other geometric/algebraic objects to the $\mathcal{I}$-graded case. We refer the reader to \cite{Jiang2022} for more detail. In this paper, we often use such objects directly without giving their definitions.
\end{rmk}

\subsection{L-manifolds}

\begin{defn}
	A $Q$-manifold is a graded manifold $\mathcal{M}$ equipped with a vector field $Q$ of degree $1$ such that $Q^2=\frac{1}{2}[Q,Q]=0$. $Q$ is called the cohomological vector field of $\mathcal{M}$.
\end{defn}

\begin{exmp}\label{QTM}
	Let $M$ be a manifold. Let $\mathcal{M}=T[1]M$. Let $(x^{\mu},\eta^{\mu})$ be a local coordinate system of $\mathcal{M}$. We define
	\begin{align*}
		Q=\eta^{\mu} \frac{\partial}{\partial x^{\mu}}.
	\end{align*}      
	The $Q$-cohomology of $\mathcal{M}$ is the de Rham cohomology of $M$.
\end{exmp}

\begin{exmp}\label{LA}
	Let $E$ be a vector bundle over $M$. Let $\mathcal{M}=E[1]$. Let $(x^{\mu},\theta^a)$ be a local coordinate system of $\mathcal{M}$. We define
	\begin{align*}
		Q=-\frac{1}{2}f^a_{bc}(x)\theta^b \theta^c \frac{\partial}{\partial \theta^a}+ \rho^{\mu}_a(x) \theta^a \frac{\partial}{\partial x^{\mu}}.        
	\end{align*}
	The condition $Q^2=0$ is equivalent to requiring that $E$ equipped with the Lie bracket $[\cdot,\cdot]:\Gamma(E)\times\Gamma(E) \rightarrow \Gamma(E)$ and the anchor map $\rho: E \rightarrow TM$ induced by $f^a_{bc}(x)$ and $\rho^{\mu}_a(x)$ is a Lie algebroid \cite{Vaintrob1997}. 
	
	In the case where $E=TM$, $f^a_{bc}(x)=0$ and $\rho_a^{\mu}(x)=\delta_a^{\mu}$, we recover Example \ref{QTM}. 
	
	As another special case, we can take $M$ to be a $G$-manifold, $E$ to be the trivial bundle $M \times \mathfrak{g}$ over $M$, $f^a_{bc}(x)$ to be the structure constants of $\mathfrak{g}$, and $\rho_a^{\mu}(x)$ to be induced from the infinitesimal action of $\mathfrak{g}$ on $M$. 
	The $Q$-cohomology of $\mathcal{M}$ is nothing but the Lie algebra cohomology of $\mathfrak{g}$ with coefficients in the $\mathfrak{g}$-module $C^{\infty}(M)$. The $Q$-closed condition of a function $S$ of degree $0$ is equivalent to the $G$-invariance condition of $S$.
\end{exmp}

$Q$-manifolds provide the most general setting for studying cohomological field theories. However, it remains to find a $Q$-closed function $S$ of even degree. $S$ will be considered as the action functional of the corresponding physical theory.

\begin{defn}
	A $P$-manifold is a graded manifold $\mathcal{M}$ equipped with a non-degenerate closed $2$-form $\omega$ of odd degree $n$. 
\end{defn}
Since $\omega$ is non-degenerate, to any $f \in \mathcal{O}(M)$ of degree $p$ one can associate a Hamiltonian vector field $v_f$ of degree $p-n$ by $\iota_{v_f}\omega + df=0$. On the other hand, $\omega$ induces a (graded) Poisson bracket $\{\cdot,\cdot\}$ on $\mathcal{O}(M)$ of degree $-n$ by setting $\{f,g\}=(-1)^{d(f)}\iota_{v_f}\iota_{v_g}\omega$. Note that $\{f,g\}=-(-1)^{d(f)}v_f(g)$. It is straightforward to verify that $v_{\{f,g\}}=[v_f,v_g]$.

\begin{defn}
	A $QP$-manifold is a $P$-manifold equipped with a Hamiltonian cohomological vector field $Q$.
\end{defn}

$QP$-manifolds were first defined in \cite{Alexandrov1997}. They provide a powerful mechanism to construct the action functional $S$ of a CohFT by setting $S$ to be the Hamiltonian function associated to $Q$.

\begin{rmk}
	The cohomological vector field $Q$ can be seen as the generator of a $(0|1)$-dimensional Lie superalgebra $\mathfrak{q}$. From this point of view, a $Q$-manifold is just a graded manifold equipped with an action of $\mathfrak{q}$. As is shown in Example \ref{LA}, the bosonic symmetries of an action functional $S$ can sometimes be fully captured by the $\mathfrak{q}$-invariance condition of $S$. If $S$ has also fermionic symmetries, $\mathfrak{q}$ should be replaced by the Lie graded algebra $L$ defined in Section 2.1. In fact, the fermionic symmetries of $S$ can be captured by the condition
	$
		\iota_a S=0.
	$
	The $Q$-closed condition of $S$ then implies directly that $\mathrm{Lie}_a S = 0$ via Cartan's magic formula $Q \iota_a + \iota_a Q = \mathrm{Lie}_a$. (With a slight abuse of notation, we use $Q$ to denote the differential of $L$.)
\end{rmk}

\begin{defn}
	An $L$-manifold is a graded manifold equipped with an $L$-action.
\end{defn}
By definition, every $L$-manifold is a $Q$-manifold. The commutative graded algebra $\mathcal{O}(M)$ of functions over $\mathcal{M}$ is an $L$-module. In particular, the $L$-module $W(\mathfrak{g}) \otimes \Omega(X)$ associated to a $G$-manifold $X$ can now be viewed as the commutative graded algebra of functions over the $L$-manifold $T[1]X \oplus \underline{\mathfrak{g}}[1] \oplus \underline{\mathfrak{g}}[2]$, where $\underline{\mathfrak{g}}$ is the trivial bundle $X \times \mathfrak{g}$. For a general $L$-manifold $\mathcal{M}$, the cohomology of $\mathcal{O}(M)_{bas}$ can be then viewed as a generalization of the equivariant cohomology.

We want to construct a basic $Q$-closed action functional $S$. 
\begin{defn}
	An $LP$-manifold is a $P$-manifold equipped with a Hamiltonian $L$-action. Namely, the fundamental vector fields generated by $\mathrm{Lie}_a$, $Q$, $\iota_a$ in $L$ are Hamiltonian vector fields. 
\end{defn}
We denote the Hamiltonian functions associated to $\mathrm{Lie}_a$ and $\iota_a$ by $L_a$ and $I_a$, respectively. 

\begin{exmp}\label{QKweil}
	Let $\mathcal{M}=\mathfrak{g}[1] \oplus \mathfrak{g}^*[2]$, where $\mathfrak{g}$ is viewed as a bundle over a point. Let $(\theta^a,\phi_b)$ be a (local) coordinate system. The canonical symplectic form on $\mathcal{M}$ takes the form $\omega = d\theta^a \wedge d \phi_a$. In this case, the Hamiltonian functions $I_a$, $L_a$ and $S$ are of degrees $2$, $3$ and $4$, respectively. We can write
	\begin{align*}
		&I_a = f_{1a}^b \phi_b + f_{2abc}\theta^b \theta^c, \\
		&L_a = h_{2ac}^{b}\phi_b\theta^c + h_{3abcd}\theta^b\theta^c\theta^d \\
		&S = g_2^{ab}\phi_a\phi_b + g_{3bc}^a \phi_a \theta^b\theta^c + g_{4abcd}\theta^a\theta^b\theta^c\theta^d,
	\end{align*}
	generally. They need to satisfy
	\begin{align*}
		\{L_a,I_b\}=f_{ab}^cI_c, \quad \{S,S\}=0,\quad \{I_a,I_b\}=0,\quad \{S,I_a\}=L_a.
	\end{align*}
	When $\mathfrak{g}$ is compact and semi-simple,\footnote{This property guarantees that $f_{ab}^c=f_{bc}^a=f_{ca}^b$ in orthonormal coordinates.} one can find a set of solutions
	\begin{align*}
		I_a = \phi_a, \quad L_a = -f^b_{ac}\phi_b\theta^c, \quad S=\frac{1}{2}g^{ab}\phi_a\phi_b - \frac{1}{2}f_{bc}^a \phi_a \theta^b\theta^c,
	\end{align*}
	where $g^{ab}$ is the Killing metric on $\mathfrak{g}$. Using $g^{ab}$ to identify $\mathfrak{g}^*$ with $\mathfrak{g}$, these solutions recover the $L$-module structure of the Weil algebra $W(\mathfrak{g})$.
\end{exmp}

The Hamiltonian function $S$ associated to $Q$ can not be basic in general, because
\begin{align*}
	\iota_a(S)=-\{I_a, S\} = -L_a.
\end{align*}
and $L_a$ cannot be $0$ when the even part of the $L$-action is non-trivial. In other words, $LP$-manifolds cannot help us to find basic $Q$-closed $S$ like the way $QP$-manifolds help one to find $Q$-closed $S$. 

\subsection{QK-manifolds}

We are particularly interested in the class of $L$-manifolds where $\mathfrak{g}$ is abelian. The $L$-structure of such graded manifold is given by $2n+1$ vector fields satisfying
\begin{align}
	&[\mathrm{Lie}_{\mu},\mathrm{Lie}_{\nu}]=0,\quad [\mathrm{Lie}_{\mu},Q]=0, \quad [\mathrm{Lie}_{\mu},\iota_{\nu}]=0, \label{qilv1}\\
	&[Q,Q]=0,\quad [Q,\iota_{\mu}]=\mathrm{Lie}_{\mu},\quad [\iota_{\mu},\iota_{\nu}]=0, \label{qilv2}
\end{align}
for $\mu,\nu = 1, \dots, n$, where $n = \dim \mathfrak{g}$.
\begin{exmp}\label{tss}
	Let $\mathcal{M} = T[1] \mathbb{R}^n$. Let $(x^{\mu},\theta^{\nu})$ be a (local) coordinate system. The $L$-structure is given by
	\begin{align}\label{tv}
		Q=\theta^{\mu} \frac{\partial}{\partial x^{\mu}}, \quad \iota_{\mu} = \frac{\partial}{\partial \theta^{\mu}}, \quad \mathrm{Lie}_{\mu} = \frac{\partial}{\partial x^{\mu}}.
	\end{align}    
	It is not hard to check that (\ref{tv}) satisfy (\ref{qilv1}) and (\ref{qilv2}). There is no interesting basic $Q$-closed function $S$ on $\mathcal{M}$. This is because $I_{\mu}(S)=0$ and $L_{\mu}(S)=0$ force $S$ to be independent of $x^{\mu}$ and $\theta^{\mu}$, hence the only possible candidates for such $S$ are constant functions.
\end{exmp}
Example \ref{tss} can be easily generalized to a non-flat case. To achieve that, we need to work in a bigraded setting instead of the graded setting. The parity function is defined by
\begin{align*}
	p: \mathbb{Z}\times \mathbb{Z} &\rightarrow \mathbb{Z}_2 \\
	(i,j) &\mapsto i+j\mod 2
\end{align*} 
Note that 
\begin{align*}
	p(d(x)d(y)),\quad p(d(x))p(d(y))
\end{align*}
are not the same in this case. As mentioned before, we use the first one as our sign convention for the bigraded setting. We say a commutative bigraded algebra $A$ is of the first (or the second) kind if $p(a)p(b)$ is set to be $p(d(a)d(b))$ (or $p(d(x))p(d(y))$), for $a, b \in A$. These two conventions can be connected by the following lemma.
\begin{lem}\label{signconvconn}
	Let $A$ be a commutative bigraded algebra of the first kind. Let $A'$ be a bigraded algebra with the same underlying bigraded vector space as $A$ and a new algebraic product $\cdot_{A'}$ defined by
	\begin{align*}
		a \cdot_{A'} b = (-1)^{j_ai_b} ab
	\end{align*}
	where $a$ is of degree $(i_a,j_a)$ and $b$ is of degree $(i_b,j_b)$. $A'$ is then a commutative bigraded algebra of the second kind. If $D$ is a derivation of $A$ of degree $(i,j)$, then $D'$ defined by
	\begin{align*}
		D'(a)=(-1)^{ji_a}D(a)
	\end{align*}
	is a derivation of $A'$.
\end{lem}
\begin{proof}
	By definition, $a \cdot_{A'} b=(-1)^{j_ai_b}ab=(-1)^{j_ai_b+i_ai_b + j_a j_b}ba=(-1)^{(i_a+j_a)(i_b+j_b)}b \cdot_{A'}a$. We need to verify that $D'$ satisfies Leibniz's rule. 
	\begin{align*}
		D'(a \cdot_{A'} b) &=(-1)^{j(i_a+i_b)}(-1)^{j_ai_b}D(ab) \\
		&=(-1)^{j(i_a+i_b)+j_ai_b}(D(a)b + (-1)^{ii_a+jj_a}aD(b)) \\
		&=(-1)^{j(i_a+i_b)+j_ai_b}((-1)^{(j+j_a)i_b}D(a) \cdot_{A'} b + (-1)^{ii_a+jj_a + j_a(i+i_b)} a \cdot_{A'} D(b)) \\
		&=D'(a) \cdot_{A'} b + (-1)^{(i+j)(i_a+j_a)}a \cdot_{A'} D'(b).
	\end{align*}
	
\end{proof}
\begin{exmp}\label{tsm}
	Consider the bigraded manifold $\mathcal{M} = (T \oplus T)[(1,1)] M$ where $M$ is an $n$-dimensional manifold. Let $(x^{\mu},\eta^{\nu},\theta^{\sigma})$ be a local coordinate system where $\eta^{\nu}$ are of degree $(1,0)$ while $\theta^{\sigma}$ are of degree $(0,1)$. Vector fields analogous to those in Example \ref{tss} are given by
	\begin{align}\label{tv3}
		Q=\theta^{\mu} \frac{\partial}{\partial x^{\mu}}, \quad K = \eta^{\mu} \frac{\partial}{\partial \theta^{\mu}}, \quad L = \eta^{\mu} \frac{\partial}{\partial x^{\mu}}.
	\end{align}
	It is easy to check that instead of (\ref{qilv1}) and (\ref{qilv2}), they satisfy
	\begin{align} 
		&Q^2=0, \quad L^2=0, \label{qkl1}\\
		&QL-LQ=0,\quad KL+LK=0,\quad QK+KQ=L.\label{qkl2}
	\end{align}
	Note that $K^2$ does not vanish and that the relations $QL-LQ=0$ and $L^2$ are not independent of the rest of (\ref{qkl1}) and (\ref{qkl2}). In fact, we have
	\begin{align*}
		QL-LQ=Q(QK+KQ)-(QK+KQ)Q=Q^2K -KQ^2=0
	\end{align*}
	by using $Q^2=0$ and $L=QK+KQ$, and
	\begin{align*}
		L^2=\frac{1}{2}(L(QK+KQ)+(QK+KQ)L)=\frac{1}{2}(QLK-KLQ-QLK+KLQ)=0
	\end{align*}
	by using $KL+LK=0$ and $L=QK+KQ$.
\end{exmp}
\begin{rmk}\label{2conv}
	If we use the second convention for $p(x)p(y)$, we will have
	\begin{align} 
		&QL+LQ=0,\quad KL-LK=0,\quad QK-KQ=L.\label{qkl2'}
	\end{align}
	instead of \eqref{qkl2}. 
\end{rmk}

\begin{defn}
	The $QK$ Lie bigraded algebra is the Lie bigraded algebra $\mathcal{K}$ spanned by $Q$ of degree $(0,1)$, $K$ of degree $(1,-1)$ and $L$ of $(1,0)$ with brackets
	\begin{align*}
		[Q,Q]=[K,K]=[L,L]=[Q,L]=[K,L]=0, \quad	[Q,K]=L.
	\end{align*}
\end{defn}

\begin{defn}
	A $QK$-manifold is a bigraded manifold $\mathcal{M}=(M,\mathcal{O})$ equipped with a $\mathcal{K}$-action.
	Equivalently, $\mathcal{M}$ is a $QK$-manifold if it has three vector fields $Q$ of degree $(0,1)$, $K$ of degree $(1,-1)$ and $L$ of degree $(1,0)$ satisfying
	\begin{align*}
		Q^2=0, \quad QK+KQ=L,\quad LK+KL=0. 
	\end{align*}
\end{defn}

\begin{defn}		
	A $QK$-algebra of length $n$ is the unital associative bigraded algebra $\mathcal{K}_n$ generated by the generators $Q$ of degree $(0,1)$, $K$ of degree $(1,-1)$ and $L$ of degree $(1,0)$ subject to the relations
	\begin{align}
		Q^2=0, \quad QK+KQ=L,\quad LK+KL=0. \label{qk}
	\end{align}
	and
	\begin{align}
		K^{n+1}=0
	\end{align}
	for some $n \in \mathbb{N}$.
\end{defn}
There is a canonical inclusion $i_{n,m}:\mathcal{K}_n \hookrightarrow \mathcal{K}_m$ for $m \geq n$. We refer to the direct limit of the system $((\mathcal{K}_n)_{n \in \mathbb{N}},(i_{n,m})_{n \leq m \in \mathbb{N}})$ as the $QK$-algebra of length $\infty$ and denote it by $\mathcal{K}_{\infty}$. Clearly, $\mathcal{K}_{\infty}$ is the unital associative bigraded algebra generated by the generators $Q$, $K$, and $L$ subject only to the relations \eqref{qk}.

\begin{rmk}
	Note that the commutative bigraded algebra of functions over a $QK$-manifold is canonically a $\mathcal{K}_{\infty}$-module (and a $\mathcal{K}_n$-module if $K^{n+1}=0$). In particular, it can be viewed as a bicomplex with horizontal differential $L$ and vertical differential $Q$. 
\end{rmk}

Let $A$ be a unital associative bigraded algebra. It can be viewed as a Lie bigraded algebra in a natural way. Let $\phi: \mathcal{K} \rightarrow A$ be a morphism of Lie bigraded algebras. 
\begin{prop}
	$\mathcal{K}_{\infty}$ is the universal enveloping algebra of $\mathcal{K}$. That is, there exists a unique morphism $\tilde{\phi}:\mathcal{K}_{\infty} \rightarrow A$ of unital associative bigraded algebras such that the following diagram commutes,
	\[
	\begin{tikzcd}
		\mathcal{K} \arrow{dr}{\phi} \arrow[hook]{r}{\iota} & \mathcal{K}_{\infty} \arrow{d}{\tilde{\phi}} \\
		& A
	\end{tikzcd}
	\]
	where $\iota: \mathcal{K} \hookrightarrow \mathcal{K}_{\infty}$ is the canonical inclusion.
\end{prop}
\begin{proof}
	This follows directly from the construction of $\mathcal{K}_{\infty}$.
\end{proof}

\begin{lem}\label{QKword}
	Every element $\alpha$ in $\mathcal{K}_n$ (or $\mathcal{K}_{\infty}$) can be uniquely written in the form
	\begin{align}\label{wordpolyk}
		\alpha = p_0(K) + p_1(K)Q + p_2(K)L + p_3(K)QL,
	\end{align}
	where $p_i(K)$ are polynomials in $K$.
\end{lem}
\begin{proof}
	Consider a word $w$ where $K$ appears to the right of $Q$, i.e.,
	\begin{align*}
		w = \cdots QK \cdots.
	\end{align*}
	One can replace $QK$ by $L-QK$ and move $L$ from the middle to the rightmost using the properties $LQ=QL$ and $LK=-KL$ to get
	\begin{align*}
		w = \cdots KQ \cdots + \cdots L.
	\end{align*}
	Repeating this procedure until there is no $K$ on the right of $Q$, one has
	\begin{align*}
		w = \sum w_Kw_{QL}
	\end{align*}
	where $w_K$ is expressed purely in $K$, $w_{QL}$ is expressed purely in $Q$ and $L$. The proof is completed by using $Q^2=L^2=0$ and $QL=LQ$.    
\end{proof}

\begin{cor}\label{PBW}
	$\mathcal{K}_{\infty} \cong \mathrm{S}[\mathcal{K}]$ as bigraded vector spaces.
\end{cor}

\begin{rmk}
	Corollary \ref{PBW} can be viewed as a special case of the Poincar\'e–Birkhoff–Witt theorem in the bigraded setting.
\end{rmk}

We prove the following lemmas for later use.
\begin{lem}\label{qpk}
	Let $p(K)$ be a polynomial in $K$, Then 
	\begin{align*}
		[Q,p(K)]=L p'(K),
	\end{align*}
	where $[\cdot,\cdot]$ is the canonical Lie bigraded bracket on $\mathcal{K}_n$ (or $\mathcal{K}_{\infty}$), $p'$ is the derivative of $p$.
\end{lem}
\begin{proof}
	Note that $[Q,\cdot]$ is a derivation of degree $(0,1)$, we have
	\begin{align*}
		[Q,K^p] &=[Q,K]K^{p-1} - K[Q,K]K^{p-2} + \cdots \\
		&= L K^{p-1} -KLK^{p-2} + \cdots \\
		&= p L K^{p-1}, 
	\end{align*}
	where we use $KL=-LK$.  
\end{proof}
Note that $\exp(K):=\sum_{p=0}^{\infty}\frac{1}{p!}K^p$ is a well-defined element in $\mathcal{K}_n$ since $K$ is nilpotent.
\begin{lem}\label{expk}
	Let $\Omega$ be a $\mathcal{K}_n$-module. We have
	\begin{align*}
		\exp(K)(\alpha \beta) = (\exp(K)\alpha)(\exp(K)\beta)
	\end{align*}
	for $\alpha,\beta \in \Omega$ with $p(d(\alpha))=0$.
\end{lem}
\begin{proof}
	Since $p(d(\alpha))=0$, we have $K(\alpha\beta)=(K\alpha)\beta+\alpha(K\beta)$. More generally, we have
	\begin{align*}
		\frac{1}{p!}K^p(\alpha \beta) &=\frac{1}{p!}\sum_{j=0}^p {p \choose  j} (K^j \alpha)(K^{p-j}\beta) \\
		&=\sum_{j=0}^p \frac{1}{j!(p-j)!} (K^j \alpha)(K^{p-j}\beta).
	\end{align*}
	It follows that
	\begin{align*}
		\exp(K)(\alpha\beta)&=\sum_{p=0}^{\infty}\sum_{j=0}^p \frac{1}{j!(p-j)!} (K^j \alpha)(K^{p-j}\beta) \\
		&=\sum_{j=0}^{\infty}\sum_{p=j}^{\infty} \frac{1}{j!(p-j)!} (K^j \alpha)(K^{p-j}\beta) \\
		&=(\sum_{j=0}^{\infty} \frac{1}{j!} K^j \alpha) (\sum_{l=0}^{\infty} \frac{1}{l!}K^{l}\beta) \\
		&=(\exp(K)\alpha)(\exp(K)\beta).
	\end{align*}
\end{proof}

\section{Cohomological Lagrangian Field Theories}

\subsection{The General Setting of CohFTs}

Let $\pi: Y \rightarrow M$ be a fiber bundle over an $n$-dimensional compact manifold $M$ with fiber $Z$ being an $m$-dimensional manifold. Let $\Gamma(Y)$ be the space of sections of $Y$. Let's consider the de Rham complex $\Omega (M \times \Gamma(Y))$ of differential forms on $M \times \Gamma(Y)$. It is a bicomplex bigraded according to the product structure of $M \times \Gamma(Y)$, i.e,
\begin{align*}
	\Omega (M \times \Gamma(Y)) = \bigoplus_{p,q} \Omega^{p,q} (M \times \Gamma(Y)).
\end{align*}
Correspondingly, the de Rham differential $d_{tot}$ on $M \times \Gamma(Y)$ breaks into two parts $d_{tot}=d + \delta$, where $d$ is the de Rham differential on $M$ and $\delta$ is the de Rham differential on $\Gamma(Y)$.

There is a canonical sub-bicomplex $\Omega_{loc}(M \times \Gamma(Y))$ \cite{Zuckerman1987}. Let $J^{\infty}Y$ be the infinite jet bundle of $Y$ over $M$. Let $\mathrm{ev}$ be the evaluation map from $M \times \Gamma(Y)$ to $J^{\infty}Y$, i.e.,
\begin{align*}
	\mathrm{ev}: M \times \Gamma(Y) &\rightarrow J^{\infty}(Y) \\
	(x,\psi) &\mapsto j^{\infty}(\psi)(x),
\end{align*}
where $j^{\infty}(\psi)(x)$ is the infinite jet prolongation of $\psi$ at $x$. The pull-back $\mathrm{ev}^* \Omega(J^{\infty}Y)$ is stable under both $d$ and $\delta$, hence a sub-bicomplex, which is called the variational bicomplex of $Y$ and denoted by $\Omega_{loc}(M \times \Gamma(Y))$. Elements in $\Omega_{loc}(M \times \Gamma(Y))$ are called local forms. $d$ and $\delta$ restricted to $\Omega_{loc}(M \times \Gamma(Y))$ are called the horizontal differential, denoted by $d_h$, and the vertical differential, denoted by $d_v$, respectively.

Let ${U}$ be an open neighborhood of $M$ such that $\pi^{-1}U \cong U \times Z$. Let ${V}$ be a coordinate chart of $Z$. $Y$ can then be covered by coordinate charts of the form $U \times V$ with coordinate functions $x^1,\dots,x^n,u^1,\dots,u^m$. Let $\mathcal{W}(U,V)$ be the set of pairs $(x,\psi)$ such that $\psi(x)$ is in $V$, it is then an open neighborhood of $M \times \Gamma(Y)$. We can explicitly define functions $x^{\mu}$ and $u^j_I$ on $\mathcal{W}(U,V)$ by setting
	\begin{align*}
		x^{\mu}(x,\psi)=x^{\mu}(x),\quad u^j_I(x,\psi)=\partial_I(u^j(\psi(x))),
	\end{align*}
where $\partial_I$ is the partial derivative in $x^{\mu}$ with respect to the multi-index $I=(\mu_1,\dots,\mu_n)$. By definition, a local function on $\mathcal{W}(U,V)$ depends only on finitely many of $x^{\mu}$ and $u^j_I$. In particular, $x^{\mu}$ and $u^j_I$ themselves are local functions. Their derivatives $dx^{\mu}$ and $\delta u^j_I$ can be viewed as local forms of degree $(1,0)$ and $(0,1)$ respectively. We can write any local $(k,l)$-form $\omega$ as a finite sum
	\begin{align*}
		\omega = f_{\mu_l,\dots,\mu_k,j_1,\dots,j_l}^{I_1,\dots,I_l} dx^{\mu_1} \wedge \dots \wedge dx^{\mu_k} \wedge \delta u^{j_1}_{I_1} \wedge \dots \wedge \delta u^{j_l}_{I_l},
	\end{align*}
where each $f^{(\dots)}_{(\dots\dots)}$ is a local function. 

In fact, the above discussion can be easily generalized to the case of a ``graded fiber bundle'' $Y$. That is, we require $Y$ to be a graded vector bundle over $Y_0$ with fiber consisting of elements of nonzero degree, where $Y_0$ is a fiber bundle over $M$ with fiber consisting of elements of degree $0$. The only subtleties of this generalization are  
\begin{enumerate}
	\item we should assign degree $(0,d(u^j))$ to the local function $u^j_I$ and degree $(0,d(u^j)+1)$ to the local form $\delta u^j_I$ where $d(u^j)$ is the degree of $u^j$ induced from the grading of $Y$;
	\item a local function should be polynomial in $u^j_I$ when $d(u^j)\neq0$.
\end{enumerate}
Local forms obtained in this way can be seen as a sheaf of commutative bigraded algebras of the second kind over $M \times \Gamma(Y_0)$. We can turn this sheaf into a sheaf of commutative bigraded algebras of the first kind by applying Lemma \ref{signconvconn}. In other words, we obtain an infinite dimensional bigraded manifold $\mathcal{M}_Y$ from $Y$. We call $\mathcal{M}_Y$ the variational bigraded manifold of $Y$. Local forms can be viewed simply as functions over $\mathcal{M}_Y$.  
\begin{rmk}
	In the physics literature, the vertical degree of a local form $\omega$ viewed as a function over $\mathcal{M}_Y$ is called as the ghost number of $\omega$. One should not confuse it with the vertical degree of $\omega$ viewed as a differential form.
\end{rmk}
The differentials $d_h$ and $d_v$ can be viewed as vector fields over $\mathcal{M}_Y$ of degree $(1,0)$ and $(0,1)$, respectively. (Note that by Lemma \ref{signconvconn}, $d_h d_v - d_v d_h =0$.) They act on $x^{\mu}$, $u^j_I$, $d x^{\mu}$ and $\delta u^j_I$ as
\begin{align*}
	&d_h(x^{\mu}) = dx^{\mu},\quad d_h(u^j_I) = u^j_{I\cup\{\mu\}} dx^{\mu}, \quad d_h(dx^{\mu}) = 0,\quad d_h(\delta u^j_I) = \delta u^j_{I\cup\{\mu\}} dx^{\mu}, \\
	&d_v(x^{\mu})=0,\quad d_v(u^j_I) = \delta u^j_I, \quad d_v(dx^{\mu})=0,\quad d_v(\delta u^j_I) = 0.
\end{align*}
We write $d_h = d_{h1} + d_{h2}$ where $d_{h1}$ and $d_{h2}$ are defined by
\begin{align*}
	&d_{h1}(x^{\mu}) = dx^{\mu},\quad d_{h1}(u^j_I) = 0, \quad d_{h1}(dx^{\mu}) = 0,\quad d_{h1}(\delta u^j_I) = 0, \\
	&d_{h2}(x^{\mu}) = 0,\quad d_{h2}(u^j_I) = u^j_{I\cup\{\mu\}} dx^{\mu}, \quad d_{h2}(dx^{\mu}) = 0,\quad d_{h2}(\delta u^j_I) = \delta u^j_{I\cup\{\mu\}} dx^{\mu}.
\end{align*}

\begin{prop}\label{qkcan}
	There is a canonical $QK$-structure on a variational bigraded manifold where $Q=d_v$, $L=d_{h2}$ and $K$ is defined as follows
	\begin{align*}
		K(x^{\mu})=0,\quad K(u^j_I) = 0,\quad K(dx^{\mu})=0, \quad K(\delta u^j_I) = u^j_{I\cup\{\mu\}} dx^{\mu}.
	\end{align*} 
\end{prop}
\begin{proof}
	$K$ is a globally well-defined vector field of degree $(1,-1)$. One can easily check that $QK+KQ=L$ and $KL+LK=0$.
\end{proof}

\begin{defn}
	A $QK_v$-manifold is a variational bigraded manifold equipped with a $\mathcal{K}$-action such that the fundamental vector field generated by $L\in \mathcal{K}$ coincides with $d_{h2}$.
\end{defn} 

A $QK_v$-structure on a variational bigraded manifold $\mathcal{M}_Y$ is hence a generalization of the canonical $QK$-structure on $\mathcal{M}_Y$. 

\begin{defn}\label{CohFT}
	A cohomological (Lagrangian) field theory is a pair $(\mathcal{M}_Y,\mathcal{L})$ where $\mathcal{M}_Y$ is a $QK_v$-manifold and $\mathcal{L}$ is a $Q$-closed function on $\mathcal{M}_Y$ of degree $(n,0)$. 
\end{defn}

Let $\Gamma_Y$ denote the graded manifold $(\Gamma(Y),\Omega)$ where $\Omega$ is the sheaf of differential forms on $\Gamma(Y)$. (We call $\Gamma$ the (graded) configuration space of the corresponding theory.) The action functional of $\mathcal{L}$ is a function $S$ of degree $0$ on $\Gamma_Y$ defined by
\begin{align*}
	S =\int_M \mathcal{L}.
\end{align*}
In most cases, the cohomological vector field $Q$ does not depend on coordinates $x^{\mu}$ and $dx^{\mu}$. Hence it can be viewed as a cohomological vector field on $\Gamma_Y$. We have $QS = Q(\int_M \mathcal{L}) = \int_M Q\mathcal{L}=0$. In other words, $\Gamma_Y$ is a $Q$-manifold equipped with a $Q$-closed function $S$.
\begin{rmk}
	From now on, every function over a variational bigraded manifold considered by us will be assumed implicitly to be independent of $x^{\mu}$. For this reason, we will often not distinguish $L=d_{h2}$ and the de Rham differential $d$.
\end{rmk}

\begin{defn}
    A pre-observable $\mathcal{O}$ is a function on $\mathcal{M}_Y$ such that $Q\mathcal{O}$ is $d$-exact and $d\mathcal{O}$ is $Q$-exact. An observable is a $Q$-closed function on $\Gamma_Y$.
\end{defn}

Let $\gamma$ be a submanifold representative of a $p$-cycle in $M$. Let $\mathcal{O}[\gamma]$ be a function on $\Gamma_Y$ defined by $\mathcal{O}[\gamma]=\int_{\gamma} \mathcal{O}$. $\mathcal{O}[\gamma]$ is $Q$-closed, hence an observable on $\Gamma_Y$. Note that the $Q$-cohomology class of $\mathcal{O}[\gamma]$ is independent of the choice of representatives of $\gamma$. In other words, we have a well-defined map
$
	H_{\bullet}(M) \rightarrow H^{\bullet}(\Gamma)
$
defined by sending $\gamma$ to $\mathcal{O}[\gamma]$.

\begin{defn}
	Let $\mathcal{O}^{(0)}$ be a $Q$-closed function of degree $(0,n)$. The descendant sequence of $\mathcal{O}^{(0)}$ is a sequence of pre-observables $\{ \mathcal{O}^{(p)}\}_{p=0}^n$ of degrees $(0,n),(1,n-1),\cdots,(n,0)$, satisfying
	\begin{align}\label{deq}
		Q \mathcal{O}^{(p)} = d \mathcal{O}^{(p-1)}
	\end{align}
    for $p=1,\dots,n$. (\ref{deq}) is called the (topological) descent equations.
\end{defn}

\begin{defn}
	Let $\mathcal{O}^{(0)}$ be a $Q$-closed function of degree $(0,n)$. The standard $K$-sequence of $\mathcal{O}^{(0)}$ is a sequence $\{ \mathcal{O}^{(p)}\}_{p=0}^n$, where 
	\begin{align*}
		\mathcal{O}^{(p)}:= \frac{1}{p!}K^p \mathcal{O}^{(0)}
	\end{align*}
	for $p=1,\dots,n$.
\end{defn}

\begin{prop}\label{kseq}
	The standard $K$-sequence is a descendant sequence.
\end{prop}
\begin{proof}
	We have $Q \mathcal{O}^{(p)}=\frac{1}{p!}Q K^p \mathcal{O}^{(0)}=\frac{1}{p!} [Q,K^p] \mathcal{O}^{(0)}=\frac{1}{(p-1)!}L K^{p-1}\mathcal{O}^{(0)}=d \mathcal{O}^{(p-1)}$ for $p \geq 1$, where we use $Q \mathcal{O}^{(0)}=0$ and Lemma \ref{qpk}.
\end{proof}

\begin{defn}
	Let $\mathcal{W}^{(q)}$ be a $Q$-closed function of degree $(q,n-q)$, $1 \leq q \leq n$. A (general) $K$-sequence of $\mathcal{O}^{(0)}$ is a sequence $\{ \mathcal{O}^{(p)}\}_{p=0}^n$, where 
	\begin{align*}
		\mathcal{O}^{(p)}:= \frac{1}{p!}K^p \mathcal{O}^{(0)} + \sum_{q=1}^p \frac{1}{(p-q)!}K^{p-q} \mathcal{W}^{(q)}
	\end{align*}
	for $p=1,\dots,n$.
\end{defn}
Likewise, one can show that
\begin{prop}\label{kgenseq}
	Every (general) $K$-sequence is a descendant sequence.
\end{prop}

\begin{rmk}
	In physics literature \cite{Sorella1998} \cite{Baulieu2005}, the vector field $K$ is known as the vector supersymmetry. A similar result as Proposition \ref{kseq} is also proved in \cite{Piguet2008}. (See Proposition 5.14 there.) 
	What we will show later is that the converse of Proposition \ref{kgenseq} is also true in a cohomological sense.
\end{rmk}

\begin{lem}\label{epp}
	Let $\{ \mathcal{O}^{(i)}\}_{i=0}^n$ be such that $\mathcal{O}^{(i)}=Q\mathcal{\rho}^{(i)} + d \mathcal{\rho}^{(i-1)}$ for $i>0$ and $\mathcal{O}^{(0)}=Q\mathcal{\rho}^{(0)}$, where $ \mathcal{\rho}^{(i)}$ is an arbitrary function of degree $(i,n-i-1)$. Then, $\{ \mathcal{O}^{i} \}_{i=0}^n$ is a solution to (\ref{deq}).
\end{lem}

\begin{proof}
	$Q\mathcal{O}^{(p)}=Q(Q\mathcal{\rho}^{(p)} + d\mathcal{\rho}^{(p-1)})=d(Q\mathcal{\rho}^{(p-1)}) = d \mathcal{O}^{(p-1)}$. 
\end{proof}

\begin{defn}\label{tpd}
	A sequence of the form in Lemma \ref{epp} is called an exact sequence. 
\end{defn}


Recall that the functions over a bigraded manifold form a bicomplex 
\begin{align*}
	\Omega = \bigoplus_{(p,q) \in [0,\dots,n]\times \mathbb{Z}} \Omega^{p,q}
\end{align*}
with commuting differentials $Q$ and $L$. Moreover, $L$ is homotopic to $0$ since $L=QK+KQ$, where $K$ is interpreted as a homotopy operator.
\[ \begin{tikzcd}
	\cdots \arrow{r}{Q} &\Omega^{p,q-1} \arrow{r}{Q} \arrow[d, shift right, swap, "0"] \arrow[d, shift left, "L"] \arrow{dl}{K} & \Omega^{p,q} \arrow{r}{Q} \arrow[d, shift right, swap, "0"] \arrow[d, shift left, "L"] \arrow{dl}{K}& \Omega^{p,q+1} \arrow{r}{Q} \arrow{r}{Q} \arrow[d, shift right, swap, "0"] \arrow[d, shift left, "L"]  \arrow{dl}{K} &  \cdots \arrow{dl}{K} \\
	\cdots \arrow{r}{Q} & \Omega^{p+1,q-1} \arrow{r}{Q} & \Omega^{p+1,q} \arrow{r}{Q} & \Omega^{p+1,q+1} \arrow{r}{Q} & \cdots
\end{tikzcd}  \] 
Let $H_{L}$ and $H_{Q}$ denote the horizontal and vertical cohomology of $\Omega$, respectively. Note that both of them are naturally bigraded. We have obtained the following result.
\begin{prop}\label{LQ=Q}
	$H^{p,q}_L(H_Q) \cong H^{p,q}_Q$ for all $0 \leq p \leq n$ and $q \in \mathbb{Z}$.
\end{prop}
Let $\Omega_{tot}$ denote the total complex of $\Omega$. Let $H_{tot}$ denote the cohomology of $\Omega_{tot}$. Let's consider the filtration $\Omega_{tot,0}^r \subset \Omega_{tot,1}^r \subset \Omega_{tot,2}^r \subset \cdots \subset \Omega_{tot}^r$ on $\Omega_{tot}^r$, where $\Omega_{tot,i}^r = \bigoplus_{\substack{p+q=r \\ p \geq n-i}} \Omega^{p,q}$. This filtration is preserved by the total differential, hence induces a filtration $H_{tot,0}^r \subset H_{tot,1}^r \subset H_{tot,2}^r \subset \cdots \subset H_{tot}^r$ on $H_{tot}^r$. Let $GH_{tot,i}^r$ denote $H_{tot,i}^r/H_{tot,i-1}^r$. We have $H_{tot}^r \cong \bigoplus_{i=0}^n GH_{tot,i}^r$.
\begin{thm}\label{mthm}
	For each $r \in \mathbb{Z}$ and $0 \leq i \leq n$, there exists a surjective map
	\begin{align} \label{surj}
		f: H_{tot,i}^r \rightarrow H_{L}^{l,r-l}(H_{Q}),
	\end{align}
	where $l=n-i$. Moreover, (\ref{surj}) induces an isomorphism
	\begin{align} \label{iso}
		GH_{tot,i}^r \cong H_{L}^{l,r-l}(H_{Q}).
	\end{align}
\end{thm}
\begin{proof}
	Let $\mathcal{O} = \sum_{l \leq p \leq n} \mathcal{O}^{p,r-p}$ be a closed element in $\Omega_{tot,i}^r$. We have $Q \mathcal{O}^{l,r-l} = 0$ and $d \mathcal{O}^{l,r-l} = Q \mathcal{O}^{l+1,r-l-1}$. Note that $\mathcal{O}^{l,r-l}$ is $Q$-exact if $\mathcal{O}$ is exact. We define $f$ to be the map induced by
	\begin{align*}
		\tilde{f}: \Omega_{tot}^r &\rightarrow \Omega^{l,r-l}\\
		\mathcal{O} &\mapsto \mathcal{O}^{l,r-l}
	\end{align*}
	$f$ is a well-defined map between cohomologies. 
	To prove the surjectivity of $f$, note that for each $\mathcal{O}^{l,r-l} \in \Omega^{l,r-l}$, the element $\sum_{p=0}^{i} \frac{1}{p!}K^p\mathcal{O}^{l,r-l} \in \tilde{f}^{-1}(\mathcal{O}^{l,r-l})$ is closed in $\Omega_{tot,i}^r$.
	The isomorphism (\ref{iso}) then follows directly from the observation that
	$\mathrm{Ker}(\tilde{f}) = \Omega_{tot,i-1}^r$.
\end{proof}

\begin{cor}\label{mcor}
	There is an isomorphism $H_{tot}^r \cong \bigoplus_{i=0}^n H_Q^{i,r-i}$ of graded modules. 
\end{cor}
\begin{proof}
	This follows directly from Proposition \ref{LQ=Q} and Theorem \ref{mthm}.
\end{proof}
\begin{rmk}
	There is a simpler proof for Corollary \ref{mcor} if we adopt the second sign convention for the bigraded setting. By Remark \ref{2conv}, $Q$ anticommutes with $L$ and commutes with $K$. The total differential is just $Q-L$. Consider the ``Mathai-Quillen map" $j=\exp(K)$ of $\Omega$. Note that the expression $\exp(K)$ is well-defined because $K$ is nilpotent in this setting. By Lemma \ref{qpk}, we have
	\begin{align*}
		j \circ Q \circ j^{-1} = \exp(K)([Q,\exp(-K)]+\exp(-K)Q) =Q-L.
	\end{align*}
    In other words, the total cohomology of $\Omega$ is equal to the $Q$-cohomology of $\Omega$. 
    
    The reason we adopt the first sign convention is just that we want to make the algebraic meaning of $K$ more transparent.
\end{rmk}

One can easily see from the proof of Theorem \ref{mthm} that Corollary \ref{mcor} is equivalent to the following statement.
\begin{thm}\label{mthmphy}
	Every descendant sequence is a $K$-sequence up to an exact sequence.
\end{thm}

Let's assume that there is a well-defined notion of integration $\int$ on $\Gamma_Y$.
The partition function $Z$ of $S$ is defined as
\begin{align*}
	Z = \int \exp(-S).
\end{align*}
The expectation value of an observable $\mathcal{O}$ is the integration
\begin{align*}
	\langle \mathcal{O} \rangle = \int \exp(-S)\mathcal{O}.
\end{align*}

In \cite{Witten1988}, assuming that $\mathcal{O}$ does not depend on the Riemannian metric $g$ of $M$, Witten observed that the expectation value $\langle \mathcal{O} \rangle$ is also independent of the choice of $g$ if the energy momentum tensor $T$ of $S$ is $Q$-exact, i.e., $T=Q(G)$ for some $G$.\footnote{CohFTs are therefore a special class of topological quantum field theories (TQFTs), though the latter do not have a constructive definition in mathematics.} More precisely, he computed
\begin{align*}
	\delta_g \langle \mathcal{O} \rangle 
	= -\int \exp(-S)T\mathcal{O} 
	= -\int \exp(-S)Q(G)\mathcal{O} 
	= -\int Q(\exp(-S)G\mathcal{O}) 
	= 0,
\end{align*}
where he used that $S$ and $\mathcal{O}$ are $Q$-closed, and that 
\begin{align}
	\mathrm{the~integration~of~a}~Q\-\mathrm{exact~function~vanishes}. \label{text}
\end{align}
	
(\ref{text}) is a bold assumption. It can be seen as an infinite dimensional analogue of Stokes' theorem. In fact, one can argue that this is indeed the case for a $QK_v$ manifold $\mathcal{M}_Y$ equipped with the canonical $QK$-structure. For such $\mathcal{M}_Y$, $Q$ is just the vertical differential $d_v$, i.e., the de Rham differential on $\Gamma_Y$. In Section 6, we will show that, in most CohFTs, $Q$ are just $d_v$ expressed in different coordinates.
\begin{rmk}
	The condition ``$T$ is $Q$-exact" is quite technical. One can consider a stronger but simpler condition, that ``$\mathcal{L}$ is $Q$-exact". There exists CohFTs where $T$ is $Q$-exact but $\mathcal{L}$ is not. For example, one has the 2-dimensional BF theory in Lorenz gauge \cite{Losev2018}. In this paper, we only consider examples with $Q$-exact Lagrangians for simplicity.
\end{rmk}

Fix a sequence $\{\gamma_i\}_{i=0}^n$ of cycles of degrees $0,1,\dots,n$ in $M$ and a descent sequence $\{ \mathcal{O}^{(i)}\}_{i=0}^n$. We can get a sequence $\{ \mathcal{O}^{(i)}[\gamma_i]\}_{i=0}^n$ of $Q$-closed observables. Obviously, $\mathcal{O}^{(i)}[\gamma_i]$ is $Q$-exact if $\{ \mathcal{O}^{(i)}\}_{i=0}^n$ is an exact sequence. Using assumption (\ref{text}), it is easy to see that $Q$-exact observables have vanishing expectation values. In other words, $K$-sequences are the only physically interesting solutions to the descent equations \eqref{deq}.

\subsection{CohFTs with Gauge Symmetries}   

Let $P$ be a principal $G$-bundle over $M$, where $G$ is a compact Lie group. The gauge symmetries are described by the automorphism group 
\begin{align*}
	\mathcal{G}:=\lbrace f: P \rightarrow P| \pi \circ f = \pi, f(pg)=f(p)g, \forall p \in P, g \in G \rbrace.
\end{align*}
Let $\mathrm{Ad}P = P \times_G G$, where $G$ acts on itself by conjugation. We have a natural identification
$
	\mathcal{G} \cong \Gamma(\mathrm{Ad}P).
$
The Lie algebra $\mathrm{Lie}(\mathcal{G})$ of $\mathcal{G}$ can then be identified as $\Gamma(\mathrm{ad}P)$, i.e., the space of sections of the adjoint bundle of $P$.

Recall that a connection $1$-form on a principal $G$-bundle $P$ is a $G$-equivariant $1$-form $A$ with values in the Lie algebra $\mathfrak{g}$ such that
$
	A(K_{\xi})=\xi,~ \xi \in \mathfrak{g},
$
where $K_{\xi}$ is the fundamental vector field generated by $\xi$ on $P$. The curvature $2$-form of $A$ is defined to be
$
	F = dA + \frac{1}{2}[A,A].
$
$F$ is a basic form, and satisfies the second Bianchi identity
$
	d_A F=0,
$
where $d_A = d + [A, \cdot]$ is the covariant derivative associated to $A$.

\begin{prop}\label{aff}
	For any principal bundle, the space of all connections $\mathcal{A}$ is an affine space modeled on $\Omega^1(\mathrm{ad}P)$. $\mathcal{A}$ has a natural $\mathcal{G}$-action, with its infinitesimal action given by
	\begin{align*}
		\mathcal{A} \times \mathrm{Lie}(\mathcal{G}) &\rightarrow T \mathcal{A} \\
		A \times \lambda &\mapsto (A,-d_A \lambda)
	\end{align*}
	where we use identifications $\mathrm{Lie}(\mathcal{G}) \cong \Omega^0(\mathrm{ad}P)$ and $T_A(\mathcal{A}) \cong \Omega^1(\mathrm{ad}P)$.
\end{prop}
For our purpose, we need to identify $\mathcal{A}$ with the space of sections of some fiber bundle over $M$. Let $P$ be a fiber bundle over $M$. Let $J^1 P$ be the first jet bundle of $P$. $J^1P$ is an affine bundle modeled on the vector bundle $T^*M \otimes_M VP$, where $VP$ is the vertical bundle over $P$ and the tensor product is taken over $M$. Let $j^1 \phi: J^1 P \rightarrow J^1 P$ denote the jet prolongation of a bundle automorphism $\phi: P \rightarrow P$. Such operations satisfy the chain rules
\begin{align*}
	&j^1(\phi_1 \circ \phi_2) = j^1(\phi_1)\circ j^1(\phi_2), \\
	&j^1(\mathrm{id}_P) = \mathrm{Id}_{J^1 P}.
\end{align*}
Thus, $J^1P$ also has a principal $G$-action. The quotient space $C=J^1 P/G$ is then an affine bundle modeled on the vector bundle $(T^*M \otimes_M VP)/G \cong T^*M \otimes \mathrm{ad}P$ over $M$.

\begin{prop}[\cite{Sardanashvili1993}]
	There exists a bijection between $\mathcal{A}$, the affine space of connection $1$-forms on $P$, and the set $\Gamma(C)$, the affine space of global sections of $C$.
\end{prop}

From now on, we will consider graded fiber bundles $Y$ of the form
\begin{align*}
	Y = C \times_M E,
\end{align*}
where $E$ is an associated bundle to $P$. (We assign degree $0$ to elements of the fiber of $C$.) Let $L_g$ denote the Lie graded algebra associated to the Lie algebra $\mathrm{Lie}(\mathcal{G})$. $L_g$ is spanned by elements $\delta_{\lambda}, \iota_{\lambda}$ and $Q_g$ for each $\lambda \in \mathrm{Lie}(\mathcal{G})$. They are of degrees $0$, $-1$, $1$, respectively, and satisfy 
\begin{align*}
	&[\delta_{\lambda_1}, \delta_{\lambda_2}]=\delta_{[\lambda_1,\lambda_2]}, \quad [\delta_{\lambda_1}, \iota_{\lambda_2}]=\iota_{[\lambda_1,\lambda_2]}, \quad [\delta_{\lambda},Q_g]=0, \\
	&[Q_g,Q_g]=0, \quad [\iota_{\lambda_1}, \iota_{\lambda_2}]=0, \quad [Q_g,\iota_{\lambda}] = \delta_{\lambda}.
\end{align*}
Note that we use the new notation $\delta_{\lambda}$ to denote the Lie derivatives.
\begin{defn}
	Let $\mathcal{M}_Y$ be a $QK_v$-manifold. An $L_g$-structure on $\mathcal{M}_Y$ is said to be compatible with the $QK_v$-structure on $\mathcal{M}_Y$ if
	\begin{enumerate}
		\item $Q_g$, $\iota_{\lambda}$ and $\delta_{\lambda}$ are of degrees $(0,1)$, $(0,-1)$, $(0,0)$, respectively;
		\item $Q_g$ coincides with $Q$, $\iota_{\lambda}$ anticommutes with $K$.
	\end{enumerate}
    $\mathcal{M}_Y$ together with the compatible $QK_v$-structure and $L_g$-structure is called a $QK_{vg}$-manifold.
\end{defn} 

\begin{defn}
	Let $\mathcal{M}_Y$ be a $QK_{vg}$-manifold. $\mathcal{M}_Y$ is said to be simple if 
	\begin{align}\label{simple}
		[\delta_{\lambda},K]=0.
	\end{align}
	It is said to be h-simple if (\ref{simple}) only hold true for horizontal functions. 
\end{defn}

\begin{lem}\label{iotaK}
	$[\iota_{\lambda}, L]=[\delta_{\lambda},K]$.
\end{lem}
\begin{proof}
	This follows from direct computations.
	\begin{align*}
		[\iota_{\lambda}, L]&=[\iota_{\lambda}, [Q,K]] \\
		&=[[\iota_{\lambda},Q],K] - [Q,[\iota_{\lambda},K]] \\
		&=[\delta_{\lambda},K],
	\end{align*}
	where we use $[\iota_{\lambda},Q] = \delta_{\lambda}$ and $[\iota_{\lambda},K]=0$.
\end{proof}

\begin{defn}\label{CohFTg}
	A cohomological (Lagrangian) gauge field theory (CohGFT) is a CohFT $(\mathcal{M}_Y,\mathcal{L})$, where $\mathcal{M}_Y$ is a $QK_{vg}$-manifold and $\mathcal{L}$ is basic with respect to the $L_g$-structure on $\mathcal{M}_Y$. The CohGFT is said to be simple (or h-simple) if $\mathcal{M}_Y$ is simple (or h-simple).
\end{defn}

$\mathcal{L}$ being basic can be seen as a generalization of the notion of gauge invariance in bosonic theories. In most cases,  $\delta_{\lambda}$, $\iota_{\lambda}$ and $Q$ do not depend on coordinates $x^{\mu}$ and $dx^{\mu}$. They then gives $\Gamma_Y$ an $L_g$-structure. The action functional $S$ is a gauge invariant $Q$-closed function on $\Gamma_Y$. 

\begin{defn}
	A gauge invariant pre-observable $\mathcal{O}$ is a basic pre-observable on $\mathcal{M}_Y$. A gauge invariant observable is a basic observable on $\Gamma_Y$.
\end{defn}

By definition, the observable $\mathcal{O}[\gamma]$ associated to a gauge invariant pre-observable $\mathcal{M}_Y$ and a cycle $\gamma$ in $M$ is a gauge invariant observable. Let $\mathcal{O}^{(0)}$ be a gauge invariant pre-observable of degree $(0,n)$. A natural question to ask is: Can we find a descendant sequence of $\mathcal{O}^{(0)}$ that is also gauge invariant? 

\begin{prop}
	The basic functions over an h-simple $QK_{vg}$-manifold $\mathcal{M}_Y$ is preserved by the $QK_v$-structure.
\end{prop}
\begin{proof}
	Let $f$ be a basic function over $\mathcal{M}_Y$. We have
	\begin{align*}
	    \iota_{\lambda} (Qf) = [\iota_{\lambda},Q]f=\delta_{\lambda} f = 0,
    \quad
    	\iota_{\lambda} (Kf) = [\iota_{\lambda},K]f=0.
    \end{align*}
    We also have
    \begin{align*}
    	\delta_{\lambda} (Qf) = [\delta_{\lambda},Q]f=0,
    \quad
    	\delta_{\lambda} (Kf) = [\delta_{\lambda},K]f=0,
    \end{align*}
    where we use the h-simple property, i.e., that $[\delta_{\lambda},K]$ vanishes for horizontal functions. Since $L=QK+KQ$, the basic functions are also preserved by $L$, hence the $QK_v$-structure.
\end{proof}
Recall that a $K$-sequence of $\mathcal{O}^{(0)}$ is specified by $Q$-closed functions $\mathcal{W}^{(1)},\cdots,\mathcal{W}^{(n)}$.
\begin{cor}
	In an h-simple CohGFT, a $K$-sequence of a gauge invariant pre-observable $\mathcal{O}^{(0)}$ is gauge invariant if $\mathcal{W}^{(1)},\cdots,\mathcal{W}^{(n)}$ are gauge invariant.
\end{cor}

Now, let's turn back to the world of homological algebras. The $QK_{vg}$ manifold $\mathcal{M}_Y$ give us a bicomplex $\Omega$ just like before. But this time we have a canonical sub-bicomplex, namely the sub-bicomplex $\Omega_{bas}$ which consists of gauge invariant elements. For an h-simple CohGFT, $\Omega_{bas}$ is stable under both $Q$, $K$ and $L$. Let $H_{tot}$ and $H_Q$ denote the total cohomology and vertical cohomology of $\Omega_{bas}$, respectively. Likewise, we have the following isomorphism 
\begin{align*}
	H_{tot}^r(\Omega_{bas}) \cong \bigoplus_{i=0}^n H_Q^{i,r-i}(\Omega_{bas}),
\end{align*}
which particularly implies that
\begin{thm}\label{mthmg}
	In an h-simple CohGFT, every gauge invariant descendant sequence is a $K$-sequence up to an exact sequence.
\end{thm}

In fact, it is necessary to consider $\Omega_{bas}$ instead of $\Omega$. This is because that there is a large class of CohGFTs with trivial $Q$-cohomologies. (For example, $\Omega$ equipped with the canonical $QK_v$-structure is $Q$-acyclic if the associated bundle $E$ to $P$ in the construction of $Y$ is a vector bundle.) Therefore, one has to restrict to $\Omega_{bas}$ to obtain nontrivial observables. 
Geometrically, $\Omega_{bas}$ determines a $QK_v$-manifold $\mathcal{M}_{bas}$ as a submanifold of the $QK_{vg}$-manifold $\mathcal{M}_Y$. The Lagrangian $\mathcal{L}$ restricted to $\mathcal{M}_{bas}$ is also a $Q$-closed function of degree $(n,0)$. Thus, $(\mathcal{M}_{bas}, \mathcal{L})$ is a CohFT. It makes more sense to consider the partition function and expectation values of observables in $(\mathcal{M}_{bas}, \mathcal{L})$, since the path integrals in $(\mathcal{M}_Y, \mathcal{L})$ always carry a redundant factor due the presence of gauge symmetries.

\section{Supersymmetric Lagrangian Field Theories}

One way to construct Coh(G)FTs is to apply ``topological twistings" to supersymmetric theories living on flat spacetimes \cite{Witten1988}. In this section, we describe this idea in detail.

\subsection{Super Poincar\'e Algebras}

The isometry group of the $n$-dimensional Minkowski space is the Lie group $\mathrm{IO}(1,n-1) \equiv \mathbb{R}^{1,n-1} \rtimes \mathrm{O}(1,n-1)$. The Lie algebra $\mathfrak{p}$ of this group is usually referred to as the Poincar\'e algebra. One can then define the Poincar\'e group to be the simply connected group uniquely determined by this Lie algebra. More precisely, it is the semi-direct product 
\begin{align*}
	\mathbb{R}^{1,n-1} \rtimes \mathrm{Spin}^0(1,n-1)
\end{align*}
where $\mathrm{Spin}^0(1,n-1)$ is the identity component of the spin group $\mathrm{Spin}(1,n-1)$. Note that $Z(\mathrm{Spin}^0(1,n-1)) \cong \mathbb{Z}_2$. The representations $\rho$ of the group $\mathrm{Spin}^0(1,n-1)$ can then be classified into two categories. We call $\rho$ a vector representation if $\mathbb{Z}_2 \subset \mathrm{ker}\ \rho$, and a spinor representation if $-1 \notin \mathrm{ker}\ \rho$. Elements in vector representations and spinor representations are called bosons and fermions, respectively. By the spin-statistics theorem in physics, bosonic fields must be commutative and fermionic fields must be anti-commutative. This is a hint that there exists a class of theories whose symmetry groups are super generalizations of the Poincar\'e groups. To define such super groups, we need the following propositions.

\begin{prop}[\cite{Deligne1999}]\label{psr}
	Let $S$ be an irreducible real spinor representation of $\mathrm{Spin}^0(1,n-1)$. 
	\begin{enumerate}
		\item The commutant $Z$ of $S$ is $\mathbb{R}$ for $n-1=0,1,7\ \mathrm{mod}\ 8$, $\mathbb{C}$ for $n-1=2,6\ \mathrm{mod}\ 8$, $\mathbb{H}$ for $n-1=3,4,5\ \mathrm{mod}\ 8$.
		\item Up to a real factor, there exists a unique symmetric morphism $\Gamma: S \otimes S \rightarrow \mathbb{R}^{1,n-1}$, and it is $Z_1$-invariant, where $Z_1$ is the group of unit elements in $Z$.
	\end{enumerate}
\end{prop}
\begin{rmk}
	For a reducible representation $S$ that is a direct sum of irreducible subrepresentations $S^{\alpha}$ with such pairings $\Gamma^{\alpha}$, one can consider $\Gamma := \sum \Gamma^{\alpha}$.
\end{rmk}
Let $S$ be a real spinor representation of $\mathrm{Spin}^0(1,n-1)$. Let $S^{\vee}$ be its dual representation. A super Poincar\'e algebra is a Lie superalgebra of the form
\begin{align*}
	\mathfrak{p}_s:= \mathfrak{p} \oplus S^{\vee}.
\end{align*}
The symmetric paring $[\cdot,\cdot]$ on the odd part of $\mathfrak{p}_s$ is given by the $\mathrm{Spin}^0(1,n-1)$-equivariant pairing $\Gamma$ in Proposition \ref{psr}. $\mathfrak{p}_s$ is said to be an $N=i$ super Poincar\'e algebra if $S$ is the direct sum of $i$ irreducible spinor representations when $n-2 \neq 0~\mathrm{mod}~4$. It is said to be an $N=(i,j)$ super Poincar\'e algebra if $S$ is the direct sum of $i$ and $j$ copies of the two inequivalent spinor representations when $n-2 =0~\mathrm{mod}~4$. The super Poincar\'e group can be then defined as the super Harish-Chandra Pair of the Poincar\'e group and the super Poincar\'e algebra.

\begin{rmk}
	Note that there is a subalgebra $\mathfrak{l}:=\mathbb{R}^{1,n-1}\oplus S^{\vee}$ of $\mathfrak{p}_s$ consisting of ``super translations''. A physical theory is usually manifestly invariant under the spin group action. Therefore, in practice, it suffices to verify that the theory is invariant under the $\mathfrak{l}$-action to show that it is supersymmetric.
\end{rmk}

The largest subgroup of the automorphism group of a super Poincar\'e algebra which fixes its underlying Poincar\'e algebra is called its $R$-symmetry group. Using the second part of Proposition \ref{psr}, one can obtain a nice classification result for $R$-symmetry groups \cite{Varadarajan2004}. See Table \ref{t_1}.
\begin{table}[!htbp]
	\centering
	\begin{tabular}{c c c c c c}
		\hline
		$n-2\ (\mathrm{mod}\ 8)$ & $1,\ 7$ & $0$ & $3,\ 5$ & $4$ & $2,\ 6$ \\
		\hline
		$\mathcal{I}_R$ & $\mathrm{SO}(N)$ & $\mathrm{SO}(N^+) \times \mathrm{SO}(N^-)$ & $\mathrm{Sp}(N)$ & $\mathrm{Sp}(N^+) \times \mathrm{Sp}(N^-)$ & $\mathrm{U}(N)$ \\
		\hline
	\end{tabular}    
	\caption{$R$-symmetry groups $\mathcal{I}_R$ for $n$-dimensional Minkowski spaces.}\label{t_1}
\end{table}

In general, one can not construct a super Poincar\'e algebra for a quadratic space of arbitrary signature, because a symmetric pairing like the one in Proposition \ref{psr} may not exist. However, this is not a problem for Euclidean spaces by applying a trick called dimensional reduction. More precisely, a super Poincar\'e algebra $\mathfrak{p}'_s$ of $\mathbb{R}^{n}$ can always be obtained from a super Poincar\'e algebra $\mathfrak{p}_s$ of $\mathbb{R}^{1,n+m}$ for $m \geq 0$ by identifying the spinor representation $S$ of $\mathrm{Spin}^0(1,n+m)$ as a spinor representation of $\mathrm{Spin}(0,n)$ via the canonical inclusion $\mathrm{Spin}(0,n) \hookrightarrow \mathrm{Spin}^0(1,n+m)$. Note that an $R$-symmetry of $\mathfrak{p}_s$ is automatically an $R$-symmetry of $\mathfrak{p}'_s$.  Therefore, one can easily find $R$-symmetry groups (or at least their subgroups) for $\mathfrak{p}'_s$ using Table \ref{t_1}.

\subsection{Twisted Super Poincar\'e Algebras and $QK$-algebras}

Let's assume that there is a well-defined supersymmetric theory living on the quadratic space $\mathbb{R}^{r,s}$. Suppose that the $R$-symmetry group $\mathcal{I}_R$ is also a symmetry of this theory. That is, we assume that the configuration space of the theory carries an action of the semi-direct product of the $R$-symmetry group and the super Poincar\'e group, under which the action functional is invariant. Since $R$-symmetry groups fix the underlying Poincar\'e algebras, we have a symmetry group of the form
\begin{align*}
	\mathrm{Spin}^0(r,s) \times \mathcal{I}_R,
\end{align*}
which is of course also a symmetry of our theory. We also assume that there exists a non-trivial group homomorphism $h_R$ from $\mathrm{Spin}^0(r,s)$ to $\mathcal{I}_R$. The term topological twisting refers to the change of the ways of embedding $\mathrm{Spin}^0(r,s) \hookrightarrow \mathrm{Spin}^0(r,s) \times \mathcal{I}_R$ through $h_R$. More precisely, we change the canonical embedding $(\mathrm{id},\mathrm{0})$ to the ``twisted" embedding $(\mathrm{id},h_R)$. 
 
\begin{exmp}[Twisting of the $N=(1,1)$ super Poincar\'e algebra of $\mathbb{R}^{4}$.]\label{t4dN11}
	In this case, we have $\mathrm{Spin}(0,4) \cong \mathrm{Sp}_+(1) \times \mathrm{Sp}_-(1)$, and a $R$ symmetry group $\mathcal{I}_R$ inherited from the the $R$ symmetry group of $N=1$ super Poincar\'e algebra of $\mathbb{R}^{1,5}$. By checking Table \ref{t_1}, we find that  $\mathcal{I}_R \cong \mathrm{Sp}(1)$. Thus, we can define the twisting homomorphism $h_R$ by setting
	\begin{align*}
		h_R: \mathrm{Sp}_+(1) \times \mathrm{Sp}_-(1) &\rightarrow \mathcal{I}_R \\
		g=(g_+,g_-) & \mapsto g_+
	\end{align*}
	The irreducible spinor representation on $\mathbb{R}^{1,5}$ gives us two inequivalent irreducible spinor representations $S^+$ and $S^-$ after applying the dimensional reduction. We have $S^+ \cong S^- \cong \mathbb{H}$ as vector spaces. The spin group acts on $S^{\pm}$ via
	\begin{align*}
		\mathrm{Spin}(0,4) \times S^{\pm} &\rightarrow S^{\pm} \\
		g=(g_+,g_-) \times s^{\pm} &\mapsto g_{\pm}s^{\pm}
	\end{align*}
	and the $R$ symmetry group acts on $S^{\pm}$ via
	\begin{align*}
		\mathcal{I}_R\times S^{\pm} &\rightarrow S^{\pm} \\
		g \times s^{\pm} &\mapsto s^{\pm}g^*
	\end{align*}
	where all the multiplications are given by quaternionic multiplications, $g^*$ denotes the conjugate of $g \in \mathbb{H}$. The two actions commutes because $\mathbb{H}$ is an associative algebra. The action of $\mathrm{Spin}(0,4) \times \mathcal{I}_R$ on $S^{\pm}$ is indeed well defined. 
	
	The new actions of $\mathrm{Spin}(0,4)$ on $S^{\pm}$ after twisting is given by
	\begin{align*}
		\mathrm{Spin}(0,4) \times S^{+} &\rightarrow S^{+} \\
		g \times s^+ &\mapsto g_+ s^+ g_+^*
	\end{align*}
	and 
	\begin{align*}
		\mathrm{Spin}(0,4) \times S^{-} &\rightarrow S^{-} \\
		g \times s^- &\mapsto g_- s^- g_+^*
	\end{align*}
	One can show that after twisting, $S^+$ becomes $ \mathbb{R} \oplus \Lambda^2_- \mathbb{R}^4$ and $S^-$ becomes $\mathbb{R}^4$ as representations of $\mathrm{Spin}(0,4)$, where $\Lambda^2_- \mathbb{R}^4$ is the vector space of anti-self-dual $2$-forms. In other words, since all bundles over $\mathbb{R}^4$ are trivial, the twisting procedure does nothing but reorganize the field components of the original theory in a different way by turning spinor fields into differential forms. The twisted supersymmetric theories can be put on a more general spacetime, because the existence of covariant differential forms put far fewer restrictions on the geometry of the spacetime manifold than the existence of covariant spinors. 
	
	Now, let's examine the twisted super Poincar\'e algebra closely. Before the twisting, the paring $[\cdot,\cdot]$ on $S^{\pm}$ is given by \cite{Baez2009}
	\begin{align*}
		[\cdot,\cdot]: (S^+ \oplus S^-) \times (S^+ \oplus S^-) &\rightarrow \mathbb{R}^4 \\
		(s^+,s^-) \times (t^+, t^-) &\mapsto t^- (s^+)^* + s^-(t^+)^*
	\end{align*}
	Identifying $\mathbb{R}$, $\Lambda^2_-\mathbb{R}^4$ with the real part and imaginary part of $\mathbb{H}$, respectively. We have
	\begin{align*}
		[\cdot,\cdot]: (\mathbb{R} \oplus \Lambda^2_- \mathbb{R}^4 \oplus \mathbb{R}^4) \times (\mathbb{R} \oplus \Lambda^2_- \mathbb{R}^4 \oplus \mathbb{R}^4) &\rightarrow \mathbb{R}^4 \\
		(\eta_1,\chi_1,\upsilon_1) \times (\eta_2,\chi_2,\upsilon_2) &\mapsto \upsilon_2(\eta_1-\chi_1) + \upsilon_1 (\eta_2-\chi_2)
	\end{align*}
	There is a sub-pairing
	\begin{align*}
		[\cdot,\cdot]: (\mathbb{R} \oplus \mathbb{R}^4) \times (\mathbb{R} \oplus \mathbb{R}^4) &\rightarrow \mathbb{R}^4  \\
		(\eta_1,\upsilon_1) \times (\eta_2,\upsilon_2) &\mapsto \upsilon_2\eta_1 + \upsilon_1\eta_2
	\end{align*}
	which gives us a subalgebra $\mathfrak{l}_t = \mathbb{R}^4 \oplus (\mathbb{R} \oplus \mathbb{R}^4)$ of the twisted super Poincar\'e algebra. Let $w_i$, $\eta$, $\upsilon_i$, $i=1,\dots,4$, be a basis of $\mathfrak{l}_t$. We have
	\begin{align}
		&[w_i,w_j]=0,\quad [w_i,\eta]=0, \quad [w_i,\upsilon_j]=0, \label{twi1}\\
		&[\eta,\eta]=0,\quad [\eta,\upsilon_i]=w_i,\quad [\upsilon_i,\upsilon_j]=0. \label{twi2}
	\end{align}
	One immediately recognizes that (\ref{twi1}) and (\ref{twi2}) are just (\ref{qilv1}) and (\ref{qilv2}) in the disguise. In other words, we have reproduced the Lie graded algebra $L$ associated to an abelian Lie algebra by twisting a super Poincar\'e algebra. The twisted supersymmetric theory can be then given naturally a $QK$-structure. Since the Lagrangian is invariant under the $\mathfrak{l}_t$-action, it is also $Q$-closed. We then obtain a $4$-dimensional CohFT. 
\end{exmp}

\begin{rmk}
	One can also work with the twisted superalgebra defined by \eqref{twi1} and \eqref{twi2} directly, and then use the standard superfield method in physics literature to construct action functionals on flat spacetimes. This idea was studied in \cite{Baulieu2008}.
\end{rmk} 

As another example, one can twist the $N=(2,2)$ super Poincar\'e algebra of $\mathbb{R}^{4}$ obtained by applying dimensional reduction to the $N=(1,1)$ super Poincar\'e algebra of $\mathbb{R}^{1,5}$. The $R$-symmetry group in this case is $\mathrm{Sp}(1) \times \mathrm{Sp}(1)$, which is isomorphic to the spin group. Therefore, there exists three different homomorphisms $h_R$ (up to automorphisms). The twisting associated to the identity one is called the geometric Langlands twisting \cite{Kapustin2007}. 

\begin{exmp}[Geometric Langlands twisting of the $N=(2,2)$ super Poincar\'e algebra of $\mathbb{R}^{4}$]\label{t4dN22}
	In this case, the odd part of the super Poincar\'e algebra is $S_l \oplus S_r$, where $S_l \cong S_r \cong S^+ \oplus S^-$, $S^{\pm} \cong \mathbb{H}$. The $R$-symmetry group acts on $S_l \oplus S_r$ via
	\begin{align*}
		(\mathcal{I}_R \cong \mathrm{Sp}_l(1) \times \mathrm{Sp}_r(1)) \times S_l \oplus S_r &\rightarrow S_l \oplus S_r \\
		(g_l,g_r) \times (s_l,s_r) &\mapsto (s_lg_l^*,s_rg_r^*).
	\end{align*}
    It is not hard to see that, after the twisting, $S_l$ becomes $\mathbb{R} \oplus \mathbb{R}^4 \oplus \Lambda_-^2\mathbb{R}^4$ and $S_r$ becomes $\mathbb{R} \oplus \mathbb{R}^4 \oplus \Lambda_+^2\mathbb{R}^4$. Again, we are only interested in the $\mathbb{R} \oplus \mathbb{R}^4$ parts of $S_l$ and $S_r$. The pairings $[\cdot,\cdot]$ on each of them are identical to the one defined in Example \ref{t4dN11}. We then obtain an abelian Lie superalgebra $\mathfrak{l}_t = \mathbb{R}^4 \oplus (\mathbb{R} \oplus \mathbb{R}^4)_l \oplus (\mathbb{R} \oplus \mathbb{R}^4)_r$. The Lie bigraded algebra $\mathcal{K}_{GL}$ associated to $\mathfrak{l}_t$ is spanned by $Q_l$ of degree $(0,1)$, $Q_r$ of degree $(0,1)$, $K_l$ of degree $(1,-1)$, $K_r$ of degree $(1,-1)$, and $L$ of degree $(1,0)$. The only non-trivial brackets between these basis elements are
    \begin{align*}
    	[Q_l, K_l]=[Q_r,K_r]=L.
    \end{align*} 
    The universal enveloping algebra $\mathcal{K}_{GL\infty}$ of $\mathcal{K}_{GL}$ is generated by $Q_l$, $Q_r$, $K_l$, $K_r$, and $L$, which are subject to the relations
    \begin{align*}
    	&Q_l^2=0, \quad Q_l K_l + K_l Q_l=L, \quad K_l L + L K_l =0, \\
    	&Q_r^2=0, \quad Q_r K_r + K_r Q_r=L, \quad K_r L + L K_r =0, \\
    	&Q_lQ_r + Q_r Q_l = 0, \quad K_l K_r - K_r K_l=0, \quad Q_l K_r + K_r Q_l = 0, \quad Q_r K_l + K_l Q_r =0.
    \end{align*}
    There exists a $\overline{T}\mathbb{RP}^1$-family of $QK$-algebras as subalgebras of $\mathcal{K}_{GL\infty}$, where $\overline{T}\mathbb{RP}^1$ is an affine bundle modelled on the tangent bundle of the $1$-dimensional projective space $\mathbb{RP}^1$. Let $(u_1,u_2,v_1,v_2) \in \mathbb{R}^4$ be such that $u_1v_1+u_2v_2=1$. We define
    \begin{align*}
    	Q_{\vec{u}}=u_1Q_l + u_2Q_r,\quad K_{\vec{v}}=v_1 K_l + v_2 K_r.
    \end{align*}
    It is straightforward to verify that
    \begin{align*}
    	Q_{\vec{u}}^2=0, \quad Q_{\vec{u}}K_{\vec{v}}+K_{\vec{v}}Q_{\vec{u}}=L, \quad K_{\vec{v}}L+LK_{\vec{v}}=0,
    \end{align*}
    where $\vec{u}=(u_1,u_2)$ and $\vec{v}=(v_1,v_2)$. Let $(u'_1,u'_2,v'_1,v'_2)$ be another element in $\mathbb{R}^4$ such that $u'_1v'_1+u'_2v'_2=1$. Obviously, $(u_1,u_2,v_1,v_2)$ and $(u'_1,u'_2,v'_1,v'_2)$ determine the same $QK$-algebra (up to a scaling factor) if there exists an $a \in \mathbb{R}/\{0\}$ such that 
    $
    	u'_1=au_1,~ u'_2=au_2,~ v'_1=v_1/a,~ v'_2=v_2/a.
    $
    On the other hand, if we fix $(u_1,u_2,v_1,v_2)$ and let $\Delta K_{\vec{u}} = -u_2 K_l + u_1 K_r$, then $Q_{\vec{u}}$, $K_{\vec{v}}+s\Delta K_{\vec{u}}$, and $L$ form a $QK$-algebra for any $s \in \mathbb{R}$.
    
    Note that there exists a natural $\mathrm{SL}(2,\mathbb{R})$-action on theses $QK$-algebras. More precisely, for 
    $g\in \mathrm{SL}(2,\mathbb{R})$, we set
    \begin{align*}
    	g Q_{\vec{u}} = Q_{g\vec{u}},\quad g K_{\vec{v}} = K_{(g^{-1})^t\vec{v}}, \quad gL=L,
    \end{align*}
    In other words, the Geometric Langlands twisting of the $N=(2,2)$ supersymmetric theory gives us a $\mathbb{RP}^1$-family of CohFTs which can be related to each other via a natural $\mathrm{SL}(2,\mathbb{R})$-action.
\end{exmp}

\begin{rmk}
	One can generalize the above examples by considering bigraded algebra generated by $2k+1$ generators: $Q_i$ of degree $(1,0)$, $K_j$ of degree $(1,-1)$, $L$ of degree $(1,0)$, $i,j=1,\cdots,k$, with non-trivial brackets being
	\begin{align*}
		[Q_i,K_i]=L, \quad i=1,\cdots,k.
	\end{align*}
    Such a bigraded algebra should be obtained from the $N=(k,k)$ super Poincar\'e algebra of $\mathbb{R}^4$. It follows that there exists a $\mathbb{RP}^{k-1}$-family of CohFTs obtained by twisting $N=(k,k)$ supersymmetric theories, which can be related to each other via a natural $\mathrm{SL}(k,\mathbb{R})$-action. For $k=1,2$, we recover Examples \ref{t4dN11} and \ref{t4dN22}. However, I do not know if it makes sense to talk about $N=(k,k)$ supersymmetric theories in physics for $k \geq 3$.
\end{rmk}

\section{Mathai-Quillen Formalism Revisited: a Generalization}

\subsection{Mathai-Quillen Formalism with Gauge Symmetries}

In this subsection, we generalize the Mathai-Quillen formalism described in Section 2 to incorporate $QK$-structures and gauge symmetries.

Let $P$ be a principal $G$-bundle over an $n$-dimensional manifold $M$. Let $\mathrm{ad} P$ denote the adjoint bundle of $P$. Let $\mathcal{A}$ denote the affine space of connection $1$-forms on $P$. Recall that $\mathcal{A}$ can be identified with $\Gamma(C)$ where $C$ is an affine bundle over $M$. Let $V$ and $W$ be two associated vector bundles to $P$. We consider the variational bigraded manifold $\mathcal{M}_Y$ associated to the graded fiber bundle
\begin{align*}
	Y=\mathrm{ad} P \times_M C \times_M V \times_M W,
\end{align*}
where the grading is defined by assigning elements of the fibers of $\mathrm{ad} P$, $C$, $V$, and $W$ degrees $1$, $0$, $-2$, and $-1$, respectively.

A bundle chart of $P$ induces a local coordinate system
\begin{align}\label{cormqm}
	(x^{\mu},~\theta^a_I,~A^a_{\mu;I},~w^i_{I},~\chi^i_I,~ dx^{\mu},~ \delta \theta^a_I,~ \delta A^a_{\mu;I},~ \delta w^i_I,~ \delta \chi^i_I)
\end{align}
for $\mathcal{M}_Y$. The degrees of the above coordinate functions are 
\[
(0,0),~(0,1),~(0,0),~(0,-2),~(0,-1),~(1,0),~(0,2),~(0,1),~(0,-1),~(0,0).
\]
For simplicity, we omit the indices of the coordinate functions and use $\phi$, $\upsilon$, $\psi$, $b$ to denote the odd coordinates $\delta \theta$, $\delta A$, $\delta w$, $\delta \chi$, respectively. There exist a family of $QK_v$-structures parameterized by $t \in \mathbb{R}$ on $\mathcal{M}_Y$. The cohomological vector field $Q$ is defined by setting
\begin{align*}
	Q \theta = \phi,~
	Q A = \upsilon,~
	Q w = \psi,~
	Q \chi = b,
\end{align*}
and the action of $Q$ on the other coordinates to be $0$. The homotopy operator $K$ is defined by setting
\begin{align*}
	K \theta = t A,~
	K \phi = d\theta - t \upsilon,~
	K \upsilon = dA,~
	K \psi = dw,~
	K b = d\chi,
\end{align*}
and the action of $K$ on the other coordinates to be $0$. $Q$ and $K$ are of degrees $(0,1)$ and $(1,-1)$, respectively. From now on, we set $t=1$.

\begin{rmk}
	The notations we adopt here need more explanation. For example, when we write $Q w = \psi$, we actually mean a family of equations $Q w^j_I = \psi^j_I$. Likewise, when we write $K \psi = dw$, we mean $K \psi^j_I = w^j_{I \cup \{\mu\}} dx^{\mu}$. The reader may question that the equations $K \theta = A$ and $K \phi = d\theta -\upsilon$ are illegal because $K$ need to be of degree $(1,-1)$. However, what we really mean by writing $A$ is $A^a_{\mu;I} dx^{\mu}$ instead of $A^a_{\mu;I}$. Likewise, we write $\upsilon$ to mean $\upsilon^a_{\mu;I} dx^{\mu}$.
\end{rmk}

By construction, $\mathcal{M}_Y$ can be equipped with an $L_g$-action. Note that the $\mathrm{Lie}(\mathcal{G})$-action on $\mathcal{A}$ is not linear. This will cause problems when we apply changes of coordinates later. Hence, we require that $\mathrm{Lie}(\mathcal{G})$ acts on $\mathcal{A}$ through the adjoint action instead. The contractions $\iota_{\lambda}$ are then defined by setting
\begin{align*}
	\iota_{\lambda} \theta = \lambda, ~\iota_{\lambda} \phi = -[\lambda,\theta],~ \iota_{\lambda} \upsilon = -[\lambda,A], ~ \iota_{\lambda} \psi = -\lambda w,~ \iota_{\lambda} b = -\lambda \chi,
\end{align*}
and its action on the other coordinates to be $0$. However, $\iota_{\lambda} K + K \iota_{\lambda} \neq 0$,\footnote{One can easily check that $(\iota_{\lambda} K + K \iota_{\lambda})\phi= \iota_{\lambda} (d \theta - \upsilon) - K [\lambda, \theta] =  d \lambda + [\lambda, A] - [\lambda, A] = d \lambda$.} i.e., the $L_g$-structure is not compatible with the $QK_v$-structure. This issue will be solved later.

We apply the Mathai-Quillen map to express the $QK$-structure in new coordinates. We have
\begin{align*}
	& Q \theta = \phi - \frac{1}{2}[\theta,\theta], \quad Q \phi = -[\theta, \phi], \\
	& Q A = \upsilon - [\theta,A], \quad Q \upsilon = - [\theta, \upsilon] + [\phi, A]\\
	& Q w = \psi - \theta w, \quad Q \psi = -\theta \psi + \phi w, \\
	& Q \chi = b - \theta \chi, \quad Q b = - \theta b + \phi \chi,
\end{align*}
as a generalization of the Kalkman differential in Remark \ref{diffutc}. We also have
\begin{align*}
	& K \theta = A, \quad K \phi = d \theta -\upsilon, \\
	& K A = 0, \quad K \upsilon = dA + [A,A], \\
	& K w = 0, \quad K \psi = dw + A w, \\
	& K \chi = 0, \quad K b = d \chi + A \chi.
\end{align*}
One can verify $Q^2=0$ and $QK+KQ=L$ by direct computations. The advantage of the new coordinates is that we have a simpler expression for $\iota_{\lambda}$, namely, we have
\begin{align*}
	\iota_{\lambda} \theta = \lambda, 
\end{align*}
and $0$ for $\iota_{\lambda}$ acting on the other coordinates. 

In order to fix the incompatibility between $K$ and $\iota_{\lambda}$, and to change the $\mathrm{Lie}(\mathcal{G})$-action on $\mathcal{A}$ back to the correct one, we apply the following change of coordinates
\begin{align*}
	\upsilon \rightarrow \upsilon - d\theta.
\end{align*}
We have now
\begin{align*}
	& Q \theta = \phi - \frac{1}{2}[\theta,\theta], \quad Q \phi = -[\theta, \phi], \\
	& Q A = \upsilon + d_A \theta, \quad Q \upsilon = - [\theta, \upsilon] - d_A \phi\\
	& Q w = \eta - \theta w, \quad Q \psi = -\theta \psi + \phi w, \\
	& Q \chi = b - \theta \chi, \quad Q b = - \theta b + \phi \chi,
\end{align*}
and
\begin{align*}
	& K \theta = A, \quad K \phi = -\upsilon, \\
	& K A = 0, \quad K \upsilon = 2F, \\
	& K w = 0, \quad K \psi = d_A w, \\
	& K \chi = 0, \quad K b = d_A \chi,
\end{align*}
where $d_A = d + A$ and $F = dA + \frac{1}{2}[A,A]$ can be interpreted as the covariant derivative and the curvature of $A$. We set $\iota_{\lambda}$ to be of the same form as before. It is then not hard to see that $\delta_{\lambda} A$ give us the correct gauge transformation of a connection $1$-form, and that $\iota_{\lambda} K + K \iota_{\lambda} =0$.\footnote{In fact, we have $\iota_{\lambda}K=K\iota_{\lambda}=0$.} 

\begin{thm}
	$\mathcal{M}_Y$ is an h-simple $QK_{vg}$ manifold.
\end{thm}
\begin{proof}
	It remains to check the h-simple property. Apply the vector fields $K \delta_{\lambda} - \delta_{\lambda} K$ to coordinate functions. The only non-vanishing one is
	\begin{align*}
		(K \delta_{\lambda} - \delta_{\lambda} K) \theta 
		= K (-[\lambda,\theta]) - \delta_{\lambda} A
		= -[\lambda, A] - d_A \lambda
		= -d \lambda.
	\end{align*}
	However, functions dependent on $\theta$ are not in the kernel of $\iota_{\lambda}$. 
\end{proof}
\begin{rmk}
	Cohomological vector fields $Q$ of the above form were first given in \cite{Ouvry1989}. It was noticed there that $Q$ recovers the scalar supersymmetry in \cite{Witten1988} by setting $\theta$ to be $0$, i.e., by restricting to the horizontal functions over $\mathcal{M}_Y$. 
\end{rmk}
It remains to find appropriate Lagrangians and observables. For the Lagrangian, we set
\begin{align}\label{labrst}
	\mathcal{L}=Q(\langle (\theta,\upsilon,\psi,\chi),(0,f_1,f_2,f_3 + b) \rangle) d\mathrm{vol},
\end{align}
where $(0,f_1,f_2,f_3)$ is a $\mathcal{G}$-equivariant vector field over $\Gamma(\mathrm{ad}P) \times \mathcal{A} \times \Gamma(V) \times \Gamma(W)$, $\langle \cdot,\cdot\rangle$ is a $\mathcal{G}$-invariant inner product on the tangent space, and $d\mathrm{vol}$ is the volume form on $M$. By construction, $\mathcal{L}$ is a $Q$-exact basic function over $\mathcal{M}_Y$. It is easy to see that $\mathcal{L}$ is an infinite dimensional generalization of \eqref{Lunifin} in the Mathai-Quillen construction of an Euler class. Since one cannot have gauge symmetries in the finite dimensional case, this generalization is essentially non-trivial.

\begin{rmk}
$f_1, f_2$ and $f_3$ are referred to as the gauge fixing functions in the physics literature. They should be carefully chosen such that $d(f_1)=-2$, $d(f_2)=0$ and $d(f_3)=0$. In this way, $\mathcal{L}$ is homogeneous of degree $(n,0)$. 
\end{rmk}

Let $\mathcal{O}^{(0)}$ be a gauge invariant pre-observable of degree $0$. It can't be $Q$-exact, otherwise, the expectation values of the corresponding observables will vanish. For simplicity, let's assume that $n$ is even. A reasonable choice is then
$
	\mathcal{O}^{(0)}=\mathrm{Tr}(\phi^{m}),
$
where $m=n/2$. Using Lemma \ref{expk}, the standard $K$-sequence of $\mathcal{O}^{(0)}$ can be found as
\begin{align}\label{kseqgauge}
	\sum_{p=0}^n \mathcal{O}^{(p)}=\exp(K)\mathcal{O}^{(0)}
	=\mathrm{Tr}(\phi_K^{m}),
\end{align}
where 
\begin{align*}
	\phi_K=\exp(K)\phi = \phi - \upsilon - F
\end{align*}
can be interpreted as the curvature $2$-form on the principal $G$-bundle $\mathcal{P}=(P\times \mathcal{A})/\mathcal{G} \rightarrow M \times \mathcal{A}/\mathcal{G}$ \cite{Baulieu1988}. In this sense, \eqref{kseqgauge} is nothing but a Chern class of $\mathcal{P}$. With a slight abuse of notation, we call $\mathcal{P}$ the universal $G$-bundle, and $\phi_K$ the universal curvature $2$-form on $\mathcal{P}$, though both of them depend apparently on the choice of $P$. Likewise, we set
\begin{align*}
	\theta_K=\exp(K)\theta = \theta + A.
\end{align*}
$\theta$ is called the universal connection $1$-form on $\mathcal{P}$. In fact, the de Rham complex of $\mathcal{P}$ is a commutative bigraded algebra of the second kind. By Lemma \ref{signconvconn}, we should reset $Q$ and $K$ to be $(-1)^pQ$ and $(-1)^qK$, where $p$ and $q$ are the horizontal degrees of the functions acted by $Q$ and $K$, respectively. The universal curvature $2$-form $\phi_K$ takes the form $\phi-\upsilon+F$ instead. Let $d_{tot}$ denote the total differential associated to $Q$ and $L$. It is not hard to show that
\begin{align*}
	\phi_K=d_{tot} \theta_K + \frac{1}{2}[\theta_K,\theta_K],
\quad
	d_{tot} \phi_K + [\theta_K, \phi_K]=0.
\end{align*}
In the next subsection, we will consider the case where $P$ is the trivial $G$-bundle over $M$. We will give a notion of Weil homomorphism in the infinite dimensional setting which sends $\theta_K$ and $\phi_K$ to connection and curvature forms on a mapping space.

\subsection{Chern-Weil Homomorphism for Mapping Spaces}

Every physical theory with non-trivial dynamics involves the notion of connections. Theories with gauge symmetries are characterized by the interpretation of connections as variables, while theories with no gauge symmetry usually have their connections fixed. One approach to construct Lagrangians for rigid CohFTs, therefore, is to start with a ``universal" CohGFT and apply a generalization of the Chern-Weil homomorphism. In this subsection, we explain this idea in detail. 

Let the graded fiber bundle $Y$ be as in the previous subsection, i.e., 
$
	Y=\mathrm{ad} P \times_M C \times_M V \times_M W.
$
Let's consider the trivial principal $G$-bundle $P$ over $M$ with $G =\mathrm{SO}(2m)$. $\mathcal{G}$, $\mathrm{Lie}(\mathcal{G})$, $\mathcal{A}$ can then be identified with $C^{\infty}(M,G)$, $C^{\infty}(M,\mathfrak{g})$ and $\Gamma(T^*M) \otimes \mathfrak{g}$, respectively. Let $V$ be a real rank $2m$ vector bundle associated to $P$ by the fundamental representation of $\mathrm{SO}(2m)$. Let $W$ be the dual bundle of $V$. $\mathcal{G}$ acts on $\Gamma(V)$ and $\Gamma(W)$ fiber-wisely. By Remark \ref{diffutc}, we need to change the $QK$-structure on $\mathcal{M}_Y$ by resetting
\begin{align*}
	Q w = \psi, \quad Q \psi = 0, \quad K w = 0, \quad K \psi = dw,
\end{align*}
The $L_g$-structure also changes correspondingly. Namely, we should redefine $\iota_{\lambda}$ by setting
\begin{align*}
	\iota_{\lambda} \psi = -\lambda w.  
\end{align*}
Moreover, we reassign degree $(0,0)$ to $w$ and degree $(0,1)$ to $\psi$. We then consider $\mathcal{L}$ of the form
\begin{align}\label{labrst1}
	\mathcal{L}=Q(\chi(w) + \langle \chi,b \rangle) d\mathrm{vol}.
\end{align}
It is easy to verify that $\mathcal{L}$ is homogeneous of degree $(n,0)$.

On the other hand, let $\Sigma$ be a $2m$-dimensional Riemannian manifold, $2m \geq n$. Let $P_{\Sigma}$ be a principal $G$-bundle over $\Sigma$ equipped with a connection $1$-form $A$. Let $Y'_0$ denote the trivial fiber bundle $M \times P_\Sigma$ over $M$. Note that $\Gamma(Y'_0)=C^{\infty}(M,P_{\Sigma})$, which can be viewed as a principal $\mathcal{G}$-bundle over the mapping space $C^{\infty}(M, \Sigma)$. Therefore, the variational bigraded manifold $\mathcal{M}_{Y'_0}$ associated to $Y'_0$ carries a canonical $L_g$-action.

\begin{lem}\label{lsimp}
    $\mathcal{M}_{Y'_0}$ equipped with the canonical $QK_v$-structure and $L_g$-structure is a simple $QK_{vg}$-manifold, hence particularly an h-simple $QK_{vg}$-manifold.
\end{lem}
\begin{proof}
	We need to show that $[K, \iota_{\lambda}] =0$ and $[K,\delta_{\lambda}]=0$ (or $[L,\iota_{\lambda}]=0$). Let $(x^{\mu}, u^j_I, dx^{\mu}, \delta u^j_I)$ be a local coordinate system. It is not hard to see that we only need to check both properties for coordinates $\delta u^j_I$. For the first one, we have
	\begin{align*}
		[K,\iota_{\lambda}] \delta u^j_I = K(\delta_{\lambda} u^j_I) + \iota_{\lambda} (u^j_{I \cup \{\mu\}} dx^{\mu}) =0,
	\end{align*}
	where we use $[Q,\iota_{\lambda}]=\delta_{\lambda}$. For the second one, we have
	\begin{align*}
		[L,\iota_{\lambda}] \delta u^j_I =  L(\delta_{\lambda} u^j_I)-(\delta_{\lambda} u^j_{I \cup \{\mu\}}) dx^{\mu}=\partial_{\mu}(\delta_{\lambda} u^j_I)dx^{\mu}-(\delta_{\lambda} u^j_{I \cup \{\mu\}}) dx^{\mu}=0.
	\end{align*}
    where we use $\delta_{\lambda} u^j_I = \partial_I (\delta_{\lambda} u^j)$.
\end{proof}

The evaluation map
\begin{align*}
	\mathrm{ev}: M \times C^{\infty}(M, P_\Sigma) \rightarrow P_\Sigma
\end{align*}
pulls back $A$ to a $\mathfrak{g}$-valued $1$-form on $M \times C^{\infty}(M, P_\Sigma)$. It decomposes into two parts: the horizontal part $A_h$ along $M$, and the vertical part $A_v$ along $C^{\infty}(M, P_\Sigma)$. Let $\nabla_h$ and $\nabla_v$ denote the covariant derivatives associated to $A_h$ and $A_v$, respectively. We can write
\begin{align*}
	\nabla_h = d + A_h, \quad \nabla_v = \delta + A_v,
\end{align*}
where $d$ is the horizontal differential and $\delta$ is the vertical differential. We have three types of curvatures:
\begin{align*}
	R_h = \nabla_h^2, \quad R_v = \nabla_v^2, \quad R_m = \nabla_v \nabla_h + \nabla_h \nabla_v, 
\end{align*} 
where the subscript $m$ of $R_m$ stands for the word "mixed". We have
\begin{align*}
	R_h = d A_h + \frac{1}{2}[A_h,A_h], \quad R_v = \delta A_v + \frac{1}{2}[A_v,A_v], \quad R_m = d A_v + \delta A_h + [A_h,A_v].
\end{align*} 
By a simple analysis of degrees, we also have the following four types of Bianchi identities.
\begin{align*}
	\nabla_h R_h = 0, \quad \nabla_v R_v = 0, \quad \nabla_v R_m + \nabla_h R_v = 0, \quad \nabla_h R_m + \nabla_v R_h =0.
\end{align*}

\begin{rmk}
	Recall that the sign convention we choose for a variational bigraded manifold is of the first kind. In particular, we should have $d \delta = \delta d$. By Lemma \ref{signconvconn}, this can be achieved by redefining $\delta$ to be $(-1)^p\delta$, where $p$ is the horizontal degree of the function acted by $\delta$. Correspondingly, the expressions for the curvature $R_m$ and the third one of the above Bianchi identities change. We have
	\begin{align*}
		 R_m = d A_v - \delta A_h + [A_h,A_v], \quad -\nabla_v R_m + \nabla_h R_v = 0.
	\end{align*}
\end{rmk}
 
$G$ can be viewed as a subgroup of $\mathcal{G}$ by identifying its element with the corresponding constant functions in $C^{\infty}(M,G)$. Thus, $A_v$, $R_v$ can be viewed as $\mathrm{Lie}(\mathcal{G})$-valued $1$-form and $2$-form of $\Omega_{loc}(M \times C^{\infty}(M,P_{\Sigma}))$, respectively. They determine maps
\begin{align*}
	\mathrm{Lie}(\mathcal{G})^* \rightarrow \Omega_{loc}^{\bullet,1}(M \times C^{\infty}(M, P_\Sigma)),\quad \mathrm{Lie}(\mathcal{G})^* \rightarrow \Omega_{loc}^{\bullet,2}(M \times C^{\infty}(M, P_\Sigma)).
\end{align*}
which induce a map
\begin{align*}
	\phi_1: \Omega_{loc}(M \times \Gamma(\mathrm{ad} P)) \rightarrow \Omega_{loc}(M \times C^{\infty}(M, P_\Sigma))
\end{align*}
sending $\theta$ and $\phi$ to $A_v$ and $R_v$, respectively. This is the usual Weil homomorphism in the infinite dimensional setting. Let $(x^{\mu},dx^{\mu}, A^a_{\mu;I}, \upsilon^a_{\mu;I})$ be a coordinate system of $\Omega_{loc}(M \times \mathcal{A})$. We can also define a connection fixing map
\begin{align*}
	\phi_2: \Omega_{loc}(M \times \mathcal{A}) \rightarrow \Omega_{loc}(M \times C^{\infty}(M, P_\Sigma))
\end{align*}
which sends $A \in \mathcal{A}$ to $A_h$ and the vertical differential $\upsilon$ of $A$ to $-R_m$. Combining $\phi_1$ and $\phi_2$, we obtain a map
\begin{align*}
	\phi_W: \Omega_{loc}(M \times \Gamma(\mathrm{ad} P) \times \mathcal{A}) \rightarrow \Omega_{loc}(M \times C^{\infty}(M, P_\Sigma)),
\end{align*}
which we refer to as the Weil homomorphism for mapping spaces. By definition, it sends the universal connection $\theta_K$ and curvature $\phi_K$ defined in the previous subsection to the connection and curvature on $M \times C^{\infty}(M, P_\Sigma)$, respectively.
\begin{thm}
	$\phi_W$ preserves the h-simple $QK_{vg}$-structure.
\end{thm}
\begin{proof}
	This follows from direct computations. Let's check first that $K \phi_W = \phi_W K$. We have
	\begin{align*}
		K (\phi_W(A)) = K A_h = 0 =\phi_W(KA), \quad K(\phi_W(\theta)) = K A_v = A_h = \phi_W(K \theta).
	\end{align*}
	Since $Q$ is just the vertical differential $\delta$ for $\mathcal{M}_{Y'_0}$, we have
	\begin{align*}
		&K R_v = K(QA_v + \frac{1}{2}[A_v,A_v])
		= L A_v - QK A_v + [A_h,A_v] 
		= dA_v - \delta A_h + [A_h,A_v] 
		= R_m, \\
		&K R_m = K(LA_v - QA_h +[A_h,A_v]) 
		= -LK A_v -L A_h - [A_h, KA_v] 
		= -2 d A_h -[A_h,A_h] 
		= -2 R_h.
	\end{align*}
	Thus,
	\begin{align*}
		K (\phi_W(\upsilon)) = -K R_m = 2 R_h =\phi_W(2 F) = \phi_W(K \upsilon), \quad K(\phi_W(\phi)) = K R_v = R_m = \phi_W(K \phi).
	\end{align*}
	The next step is to check $Q \phi_W = \phi_W Q$. We have
		\begin{align*}
		&\phi_W(Q\theta) = \phi_W(\phi) - \frac{1}{2}\phi_W([\theta,\theta]) 
		= R_v - [A_v,A_v]
		= \delta A_v 
		= Q(\phi_W(\theta)),\\
    	&\phi_W(Q\phi) = \phi_W(-[\theta,\phi]) 
    	= - [A_v,R_v]
    	= \delta R_v
    	= Q(\phi_W(\phi)),
    \end{align*}
    where we use the Bianchi identity $\nabla_vR_v=0$. We also have
	\begin{align*}
		&\phi_W (QA) = \phi_W(\upsilon) + \phi_W(d_A \theta) 
		= -R_m + dA_v + [A_h,A_v] 
		= -\delta A_h 
		= Q(\phi_W(A)),\\
		&\phi_W (Q\upsilon) = \phi_W(-[\theta,\upsilon]) - \phi_W(d_A \phi) 
		= [A_v,R_m] - d R_v - [A_h,R_v]
		= -\delta R_m
		= Q(\phi_W(\upsilon)),
	\end{align*}
	where we use the Bianchi identity $-\nabla_v R_m + \nabla_h R_v=0$. It follows that	
	\begin{align*}
		L \phi_W = \phi_W L.
	\end{align*}
	To prove that $\phi_W$ preserves the $L_g$-structure, it suffices to check  $\phi_W \iota_{\lambda} = \iota_{\lambda} \phi_W$. By construction, we have
	\begin{align*}
		\iota_{\lambda}\phi_W(\theta)=\iota_{\lambda} A_v = \lambda,~\iota_{\lambda} \phi_W(\phi)=\iota_{\lambda} R_v =0,~ \iota_{\lambda}\phi_W(A)=\iota_{\lambda} A_h = 0.
	\end{align*}
    It remains to show that $\iota_{\lambda} \phi_W(\upsilon) = 0$. In fact, we can check that
    \begin{align*}
    	\iota_{\lambda} R_m &= \iota_{\lambda} (L A_v - Q A_h +[A_h, A_v]) \\
    	&= [\iota_{\lambda}, L] A_v + L \lambda - [\iota_{\lambda}, Q]  A_h + Q \iota_{\lambda} A_h + [A_h, \lambda] \\
    	&=[\delta_{\lambda}, K] A_v - \delta_{\lambda} A_h + [A_h, \lambda] \\
    	&= -K (\delta_{\lambda} A_v) + [A_h, \lambda] \\
    	&= K ([\lambda, A_v] ) + [A_h, \lambda] \\
    	&= 0,	
    \end{align*}
    where we use $[\iota_{\lambda}, L]=[\delta_{\lambda}, K]$.
\end{proof}
Let $Y'=Y'_0 \times_M V \times_M W$. We can extend $\phi_W$ naturally to a map
\begin{align*}
	\Omega_{loc}(M \times \Gamma(Y)) \rightarrow \Omega_{loc}(M \times \Gamma(Y')),
\end{align*}
which is denoted again by $\phi_W$ with a slight abuse of notation. There exists a natural map 
\begin{align*}
    \Gamma(Y'_0) \cong C^{\infty}(M,P_{\Sigma}) &\rightarrow \Gamma(T^*M) \otimes \mathfrak{g} \cong \Gamma(Y_0) \\
    f &\mapsto A(Tf),
\end{align*}
where we identify the tangent map $Tf$ of $f$ as a section of $T^*M \otimes f^* TP_{\Sigma}$, and identify $A$ as a map $\Gamma(TP_{\Sigma}) \rightarrow \mathfrak{g}$. This map together with $\phi_W$ determines a morphism $\mathcal{M}_{Y'} \rightarrow \mathcal{M}_Y$ of h-simple $QK_{vg}$-manifolds.

From now on, we choose $P_{\Sigma}$ to be the frame bundle of $\Sigma$ and $A$ to be the Levi-Civita connection. $(\Omega_{loc}(M \times \Gamma(Y'))_{bas}$ can be identified with $\Omega_{loc}(M \times \Gamma(Y_{\Sigma}))$, where $Y_{\Sigma}$ is the trivial graded fiber bundle $M \times (T\Sigma \times_{\Sigma} T^*\Sigma)$ over $M$, with $(Y_{\Sigma})_0$ being the trivial bundle $M \times \Sigma$ over $M$. $\phi_W$ induces a homomorphism 
\begin{align*}
	\phi_{CW}: \Omega_{loc}(M\times \Gamma(Y))_{bas} \rightarrow	\Omega_{loc}(M \times \Gamma(Y_{\Sigma})),
\end{align*}
which we refer to as the Chern-Weil homomorphism for mapping spaces. Let $(x, u, w, \chi, dx, \delta u, \psi, b)$ be a local coordinate system of $\mathcal{M}_{Y_{\Sigma}}$. The $QK$-structure on $\mathcal{M}_{Y_{\Sigma}}$ is given by
\begin{align*}
	& Qu = \delta u, \quad  Q \delta u =0, \quad Q w = \psi, \quad Q \psi = 0, \quad Q \chi = b - A_v \chi, \quad Q b = - A_v b + R_v \chi, \\
	& Ku = 0, \quad K \delta u = du, \quad K w = 0, \quad K \psi = dw, \quad K \chi = 0, \quad K b = d \chi + A_h \chi.
\end{align*}
To define a CohFT, we simply set the Lagrangian to be
\begin{align}\label{labrst2}
	\mathcal{L}=Q(\chi(f) + \langle \chi,b \rangle) d\mathrm{vol},
\end{align}
where $\langle \cdot,\cdot\rangle$ is the inner product on $C^{\infty}(M,T^*\Sigma)$ induced by the Riemannian metric on $\Sigma$, $f$ is a vector field over $C^{\infty}(M,\Sigma)$, and $d\mathrm{vol}$ is the volume form on $M$. \eqref{labrst2} can be seen as the pullback through $f$ of the image of \eqref{labrst1} under $\phi_{CW}$.

\begin{rmk}
	More generally, one can take $W$ in the construction of $\mathcal{M}_Y$ and $\mathcal{M}_{Y'}$ to be the tensor product of the dual bundle of $V$ and $\Lambda^p T^*M$, and $f$ to be a section of the bundle $\Lambda^p T^*M \times C^\infty(M, T\Sigma)$ over $M \times C^{\infty}(M, \Sigma)$.
\end{rmk}
\begin{rmk}
	Lagrangians of the form \eqref{labrst2} can be found in \cite{Baulieu1989,Blau1993}
\end{rmk}

For pre-observables of the CohFT, consider the map 
\begin{align*}
	\mathrm{e}: M \times \Gamma(Y_{\Sigma}) \cong M \times C^{\infty}(M, T\Sigma) \times C^{\infty}(M, T^* \Sigma) &\rightarrow \Sigma \\
	(x, f_1, f_2) &\mapsto \pi(f_1(x)),	
\end{align*}
where $\pi: T\Sigma \rightarrow \Sigma$ is the canonical projection. Let $\alpha$ be a closed $n$-form on $\Sigma$. Let
$
\mathcal{O}=\mathrm{e}^*\alpha.
$
$\mathcal{O}$ can be decomposed as $\mathcal{O} = \sum_{p=0}^n\mathcal{O}^{(p)}$, where $\mathcal{O}^{(p)}$ is of horizontal degree $p$. Locally, $\alpha$ can be written as $\alpha_{i_1,\cdots,i_n}du^{i_1} \wedge \cdots \wedge du^{i_n}$. We then have
\begin{align*}
	\mathcal{O}^{(p)}={n \choose p}\alpha_{i_1,\cdots,i_n}d_hu^{i_1} \wedge \cdots \wedge d_h u^{i_p} \wedge \delta u^{i_{p+1}} \wedge \cdots \wedge \delta u^{i_n}.
\end{align*}
One can check that $\mathcal{O}^{(p)} = \frac{1}{p} K \mathcal{O}^{(p-1)}$. In other words, $\{\mathcal{O}^{(p)}\}_{p=0}^n$ is the standard $K$-sequence of $\mathcal{O}^{(0)}$.

\section{Examples}

In this section, various Coh(G)FTs living on spacetimes of dimensions $n\leq 4$ proposed by physicists are constructed using the framework developed in Section 6. Further examples like topological F theory and Kapustin-Witten theory can also be incorporated into this framework with ease.

\begin{exmp}[Topological quantum mechanics]
The geometric setting is specified by the following data.
\begin{enumerate}
	\item $M$ is the real line $\mathbb{R}$;
	\item $\Sigma$ is a Riemannian manifold equipped with a Morse function $h$.
\end{enumerate}
The Lagrangian is specified by the gauge fixing function
\begin{align*}
	f=\frac{du}{dt} + \mathrm{grad} h,
\end{align*} 
where $t$ is the parameter of $\mathbb{R}$. The pre-observables are specified by the $1$-form $\alpha=dh$ on $\Sigma$. We have
\begin{align*}
	&\mathcal{O}^{(0)}=\partial_i h \delta u^i, \\ &\mathcal{O}^{(1)}=\partial_i h u_t^i dt.
\end{align*}
\end{exmp}

\begin{rmk}
	In dimension $1$, there is no spinor or $R$-symmetry, hence no topological twisting. The topological quantum mechanics is just the $N=2$ supersymmetric quantum mechanics.
\end{rmk}

\begin{exmp}[Topological sigma model]
	The geometric setting is specified by the following data.
	\begin{enumerate}
		\item $(M,j)$ is a Riemann surface;
		\item $(\Sigma,\omega, J)$ is a Kähler manifold.
	\end{enumerate}
	The Lagrangian is specified by the gauge fixing function
	\begin{align*}
		f=\bar{\partial}_J u,
	\end{align*} 
	where $\bar{\partial}_J u = \frac{1}{2}(D u + J \circ Du \circ j)$, $Du$ is the total differential of $u$. Note that $f$ is a section of the bundle $T^*M \times C^\infty(M, T\Sigma)$ over $M \times C^{\infty}(M, \Sigma)$. The pre-observables are specified by $\alpha=\omega$, the symplectic form on $\Sigma$. We have
	\begin{align*}
		&\mathcal{O}^{(0)}=\omega_{i_1i_2} \delta u^{i_1} \delta u^{i_2}, \\
		&\mathcal{O}^{(1)}=2\omega_{i_1i_2} u_{\mu}^{i_1} dx^{\mu} \delta u^{i_2}, \\
		&\mathcal{O}^{(2)}=\omega_{i_1i_2} u_{\mu}^{i_1} u_{\nu}^{i_2} dx^{\mu} dx^{\nu}. 
	\end{align*}
\end{exmp}
\begin{rmk}
	The topological sigma model can also be obtained by twisting the $N=2$ supersymmetric non-linear sigma model.
\end{rmk}

\begin{exmp}[Topological M theory]
	The geometric setting is specified by the following data.
	\begin{enumerate}
		\item $M$ is a $3$-dimensional manifold;
		\item $(\Sigma, \Phi)$ is a $G_2$-manifold.
	\end{enumerate}
	The Lagrangian is specified by the gauge fixing function
	\begin{align*}
		f=[Du,Du,Du],
	\end{align*} 
	where $[\cdot,\cdot,\cdot]$ is the associator bracket of $\Phi$. Note that $f$ is a section of the bundle $\Lambda^3 T^*M \times C^\infty(M, T\Sigma)$ over $M \times C^{\infty}(M, \Sigma)$. The pre-observables are specified by $\alpha=\Phi$, the closed $G_2$-structure on $\Sigma$. We have
	\begin{align*}
		&\mathcal{O}^{(0)}=\Phi_{i_1i_2i_3} \delta u^{i_1} \delta u^{i_2} \delta u^{i_3}, \\
		&\mathcal{O}^{(1)}=3\Phi_{i_1i_2i_3} u_{\mu}^{i_1} dx^{\mu} \delta u^{i_2} \delta u^{i_3}, \\
		&\mathcal{O}^{(2)}=3\Phi_{i_1i_2i_3} u_{\mu}^{i_1} u_{\nu}^{i_2} dx^{\mu} dx^{\nu} \delta u^{i_3}, \\
		&\mathcal{O}^{(3)}=\Phi_{i_1i_2i_3} u_{\mu}^{i_1} u_{\nu}^{i_2} u_{\sigma}^{i_3} dx^{\mu} dx^{\nu} dx^{\sigma}. 
	\end{align*}
\end{exmp}


\begin{exmp}[Topological Yang-Mills theory]
	The geometric setting is specified by the following data.
	\begin{enumerate}
		\item $P$ is a principal $\mathrm{SU}(2)$-bundle;
		\item $V=\mathrm{ad}P$, $W=\mathrm{ad}P \otimes \Lambda^2_-(T^*M)$.
	\end{enumerate}
	The Lagrangian is specified by the gauge fixing functions
	\begin{align*}
		f_1=d_A w, \quad f_2=[\phi,w], \quad f_3 =F_-,
	\end{align*} 
	where $F_-$ is the anti-self-dual part of the curvature $F$. Note that $f_1$ is of degree $-2$, $f_2$ is of degree $0$, and $f_3$ is of degree $0$. The pre-observables are determined by  $\mathcal{O}^{(0)}=\mathrm{Tr}(\phi^2)$. We have
	\begin{align*}
		&\mathcal{O}^{(0)}=\mathrm{Tr}(\phi^2), \\
		&\mathcal{O}^{(1)}=-2\mathrm{Tr}(\phi \upsilon), \\
		&\mathcal{O}^{(2)}=\mathrm{Tr}(\upsilon \wedge \upsilon - 2 \phi F), \\ 
		&\mathcal{O}^{(3)}=2\mathrm{Tr}(\upsilon \wedge F),\\
		&\mathcal{O}^{(4)}=\mathrm{Tr}(F \wedge F).
	\end{align*}
\end{exmp}

\begin{rmk}
	The topological Yang-Mills theory can also be obtained by twisting the $N=(1,1)$ supersymmetric Yang-Mills theory. The $QK$-structure obtained from the twisting is more complicated than the $QK$-structure defined in Section 6 \cite{Labastida2005}. More precisely, since $\chi$ and $b$ are $\mathfrak{su}(2)$-valued (anti-self-dual) $2$-forms, there exists a family of $QK$-structures parameterized by $(r,s,t) \in \mathbb{R}^3$ by setting
	\begin{align*}
		&Q \theta = \phi,~
		Q A = \upsilon,~
		Q w = \psi,~
		Q \chi = b + r F_-,~
		Q b = -r (d_A \upsilon)_-,\\
		&K A = s \chi,~
		K \theta = t A,~
		K \phi = d\theta - t \upsilon,~
		K \upsilon = dA - s (b + rF_-),~
		K \psi = dw,~
		K b = d\chi + rs d_A \chi,
	\end{align*}
	and the action of $Q$ and $K$ on the other coordinates to be $0$. Likewise, we set $t=1$. After applying the change of coordinates induced by the Mathai-Quillen map, we get
	\begin{align*}
		& Q \theta = \phi - \frac{1}{2}[\theta,\theta], \quad Q \phi = -[\theta, \phi], \\
		& Q A = \upsilon + d_A \theta, \quad Q \upsilon = - [\theta, \upsilon] - d_A \phi, \\
		& Q w = \psi - [\theta, w], \quad Q \psi = -[\theta, \psi] + [\phi, w], \\
		& Q \chi = rF_- + b - [\theta, \chi], \quad Q b = -r(d_A \upsilon)_- - [\theta, b] + [\phi, \chi],
	\end{align*}
	and
	\begin{align*}
		& K \theta = A, \quad K \phi = - \upsilon, \\
		& K A = s \chi, \quad K \upsilon = 2 F - s(b + rF_-), \\
		& K w = 0, \quad K \psi = d_A w, \\
		& K \chi = 0, \quad K b = (1+rs)d_A \chi.
	\end{align*}
	We then have
	\begin{align*}
		\theta_K = \theta + A + \frac{s}{2}\chi, \quad \phi_K = \phi - \upsilon - F + \frac{s}{2} (b+rF_-) + \frac{s}{2}d_A \chi + \frac{s^2}{8}[\chi,\chi].
	\end{align*}
    Let's set $r=0$ for simplicity. The standard $K$-sequence of $\mathcal{O}^{(0)}=\mathrm{Tr}(\phi^2)$ takes the form
    \begin{align*}
    	&\mathcal{O}^{(0)}=\mathrm{Tr}(\phi^2), \\
    	&\mathcal{O}^{(1)}=-2\mathrm{Tr}(\phi \upsilon), \\
    	&\mathcal{O}^{(2)}=\mathrm{Tr}(\upsilon \wedge \upsilon + \phi (sb-2F)), \\ 
    	&\mathcal{O}^{(3)}=\mathrm{Tr}(s\phi d_A \chi -\upsilon \wedge (sb-2F),\\
    	&\mathcal{O}^{(4)}=\mathrm{Tr}((sb/2-F) \wedge (sb/2-F) - s \upsilon \wedge d_A \chi + s^2\phi [\chi,\chi]/4).
    \end{align*}
    Let $K_0$ denote the vector symmetry in the special case of $s=0$, i.e., the vector symmetry defined in Section 6. One can easily check the above standard $K$-sequence is nothing but the general $K_0$-sequence specified by 
    \begin{align*}
    	\mathcal{W}^{(1)}=0, \quad 
    	\mathcal{W}^{(2)}=s\mathrm{Tr}(\phi b), \quad 
    	\mathcal{W}^{(3)}=0, \quad 
    	\mathcal{W}^{(4)}=-\frac{s^2}{4}\mathrm{Tr}(b \wedge b +  \phi [\chi,\chi]).
    \end{align*}
    Note that both $\mathcal{W}^{(2)}$ and $\mathcal{W}^{(4)}$ are gauge invariant and $Q$-exact. In fact, we have
    \begin{align*}
    	\mathcal{W}^{(2)}= s Q (\mathrm{Tr}(\phi \chi)), \quad \mathcal{W}^{(4)}=-\frac{s^2}{4}Q(\mathrm{Tr}(b \wedge \chi)).
    \end{align*}
    The standard $K$-sequence is equivalent to the standard $K_0$-sequence up to an exact sequence.
\end{rmk}   

\section*{Acknowledgement}

The author would like to thank Alessandro Tanzini for pointing out that Proposition \ref{kseq} was already proved in \cite{Piguet2008}. He would like to thank Jürgen Jost and Ruijun Wu for various discussions. He would also like to thank an anonymous referee for inspiring comments. This work was supported by the International Max Planck Research School Mathematics in the Sciences.

\end{document}